\documentclass[aps,pra,twocolumn,preprintnumbers,nofootinbib,superscriptaddress,amsmath,amssymb,longbibliography,]{revtex4-2}

\usepackage[utf8]{inputenc}
\usepackage[T1]{fontenc}
\usepackage{amsmath}
\usepackage{amssymb}
\usepackage{mathtools}
\usepackage{orcidlink}
\usepackage{braket}
\usepackage{multirow}
\usepackage{amsfonts}
\usepackage{graphicx}
\usepackage{float}
\usepackage{physics}
\usepackage{quantikz}
\usepackage{float}    
\usepackage{xcolor}
\usepackage{bm}
\usepackage{xcolor}
\usepackage{placeins}
\usepackage[normalem]{ulem}
\usepackage{hyperref}

\begin{document}

\title{Finite-size resource scaling for learning quantum phase transitions \\ with fidelity-based support vector machines}

\author{Aaqib Ali~\orcidlink{0009-0005-1357-1296}}
\affiliation{Dipartimento Interateneo di Fisica, Università di Bari, 70126 Bari, Italy}
\affiliation{INFN, Sezione di Bari, 70126 Bari, Italy}
\affiliation{International Centre for Theory of Quantum Technologies, University of Gda\'nsk, 80-308 Gda\'nsk, Poland}

\author{Giovanni Scala~\orcidlink{0000-0003-2685-0946}}
\affiliation{Dipartimento Interateneo di Fisica, Politecnico di Bari, 70126 Bari, Italy}
\affiliation{INFN, Sezione di Bari, 70126 Bari, Italy}

\author{Cosmo Lupo~\orcidlink{0000-0002-5227-4009}}
\affiliation{Dipartimento Interateneo di Fisica, Politecnico di Bari, 70126 Bari, Italy}
\affiliation{Dipartimento Interateneo di Fisica, Università di Bari, 70126 Bari, Italy}
\affiliation{INFN, Sezione di Bari, 70126 Bari, Italy}

\author{Antonio Mandarino~\orcidlink{0000-0003-3745-5204}}
\affiliation{International Centre for Theory of Quantum Technologies, University of Gda\'nsk, ul. prof. Marii Janion, 4 80-308 Gda\'nsk, Poland}

\begin{abstract}
Quantum kernels offer a valid procedure for learning quantum phase transitions on quantum processing devices, 
yet issues on the scalability of the learning strategy in connection with the symmetry of the critical model have not been clarified.
We derive a link between model symmetry and fidelity-kernel resource scaling.
We quantify the measurement resources required to estimate fidelity-based quantum kernels for many-body ground states while preserving the structure of the resulting Gram matrix under finite-shot sampling. 
Crucially, we show that increasing symmetry in the underlying spin model systematically amplifies these shot requirements.
Moving from the $\mathbb{Z}_2$-symmetric Ising/XY regimes to the $U(1)$-symmetric XX (and XXZ) regimes leads to stronger kernel concentration and therefore substantially larger shot costs under the same bounds.
We consider a tunable one-dimensional spin-$\tfrac{1}{2}$ Hamiltonian spanning the transverse-field Ising, XY, XX, and XXZ limits, and define the kernel as the ground-state fidelity. Kernel entries are estimated using a SWAP-test estimator with $S$ shots, and we adapt the ensemble spread and concentration-avoidance shot bounds to obtain practical shot requirements in terms of the interquartile range of kernel values and a representative kernel magnitude. 
For the free-fermion XY/XX family, we use the closed-form Bogoliubov-angle fidelity, while for the interacting XXZ chain we compute fidelities by exact diagonalization and benchmark shot-noise effects. 
Our symmetry-aware bounds provide a pragmatic procedure for physics-informed quantum machine learning.

\end{abstract}

\maketitle

\section{Introduction}

Spin chains are paradigmatic models in which spin-$\tfrac{1}{2}$ particles interact along a one-dimensional lattice. Despite their apparent simplicity, they capture a broad range of many-body emerging phenomena, to name a few formation of magnetic domains in materials, generation of multipartite quantum correlations, and even topological order~\cite{Sachdev2011QPT,Giamarchi2004Q1D,AmicoFazioOsterlohVedral2008RMP,Wen2017TopoZoo,LatorreRiera2009ShortReview}. In all these phenomena, \emph{quantum phase transitions} (QPTs) play a crucial role. They are zero-temperature transitions between distinct ground-state phases of a many-body system, driven not by thermal fluctuations but by quantum fluctuations~\cite{bethe1931theorie,pfeuty1970one,vidal2003entanglement,affleck1987rigorous}. QPTs reveal how quantum fluctuations induce abrupt changes in the structure of the ground state, unveiling universal behaviour and critical phenomena that are fundamental to understanding quantum many-body physics. 
Although QPTs are strictly defined for $T{=}0$ phenomena, their signatures have been observed in analog quantum simulation at ultra-low temperatures, typically in the nK regime~\cite{Bloch2008RMPUltracold, browaeys2020many, Zhang2012ScienceQC,Gegenwart2008NatPhysQC,Coldea2010ScienceE8}.
Despite long research efforts, QPTs are one of the most challenging problems in modern physics, 
since the scaling of correlations close to the critical points plays a pivotal role in lattice field theories as well as in non-equilibrium processes. 
As a result, considerable effort has focused on computational approaches to characterize critical points and finite-size scaling. 
The current effort focuses on utilizing methods rooted in machine learning and data-driven approaches. 
The unifying concept stems from the common statistical approach of learning strategies and systems exhibiting phase transitions. 
Instead of modeling all microscopic degrees of freedom, one can learn a decision rule from data using similarity in an appropriate feature space.

Machine learning (ML) has emerged as a powerful framework for detecting and characterizing quantum phase transitions (QPTs) across many-body systems ~\cite{qpt_wittek, qpt_melko, qpt_sarma, qpt_fanchini,CarrasquillaMelko2017}. Existing approaches broadly fall into two complementary categories. Unsupervised methods, such as clustering techniques, principal-component analysis, confusion schemes, and Fisher-information-based learning ~\cite{Kasatkin2024, Wang2016UML,Wetzel2017PRE,vanNieuwenburg2017Confusion}, aim to identify phase boundaries directly from the data without requiring prior labeling, making them particularly attractive when the phase diagram is unknown. By contrast, supervised methods learn a classification rule from representative examples belonging to different phases. Rather than discovering phases from scratch, they exploit labeled data to construct predictive models that can efficiently recognize new states and quantify decision boundaries. The present work belongs to this second class and addresses the practical implementation of supervised quantum learning on quantum hardware.

A particularly powerful supervised learning strategy is provided by support vector machines (SVMs), which have been successfully applied to both classical statistical mechanics~\cite{Giannetti_2019}  and quantum many-body systems. Instead of relying on explicit order parameters or handcrafted physical observables, SVMs learn the relevant phase structure by comparing states through a kernel function that defines a suitable feature space. In the quantum setting, this feature space is naturally constructed from the geometry of many-body ground states through fidelity-based quantum kernels. Our objective is therefore not to develop a new algorithm for unsupervised phase discovery, but to investigate the feasibility of implementing fidelity-based supervised learning on quantum processors. In this context, the dominant challenge is the estimation of the quantum kernel itself, motivating the resource analysis carried out in the remainder of this work.

Quantum support vector machines (QSVMs)~\cite{Mengoni201965} offer the ability to explore the transitions encoding many-body ground states into feature spaces where kernel evaluations yield quantum-state overlaps~\cite{SanchoLorente2022PRA,Wu2023Quantum,wang2025}. 
While the capability of fidelity kernels to identify second-order phase transitions has already been established~\cite{SanchoLorente2022PRA}, 
their finite-resource implementation on quantum hardware and how the symmetry of the underlying model affects measurement costs remained largely unexplored. The present work addresses this question.
We adopt fidelity-based quantum kernels for compelling reasons that recently emerged in the literature. It has been proven that they can offer complexity advantages over classical kernels~\cite{SchuldKilloran2019PRL,LiuArunachalamTemme2021,HuangEtAl2021}, and nowadays experimental platforms are suitable to realize them ~\cite{Havlicek2019Nature}. 
Moreover, theoretical insight on the successful classification of symmetry-protected topological and symmetry-broken phases has been reported in ~\cite{SanchoLorente2022PRA,Wu2023Quantum}, and expanded, taking into account the kernel concentration and sample complexity~\cite{Thanasilp2024}. 

\begin{figure*}
    \centering
    \includegraphics[width=\textwidth]{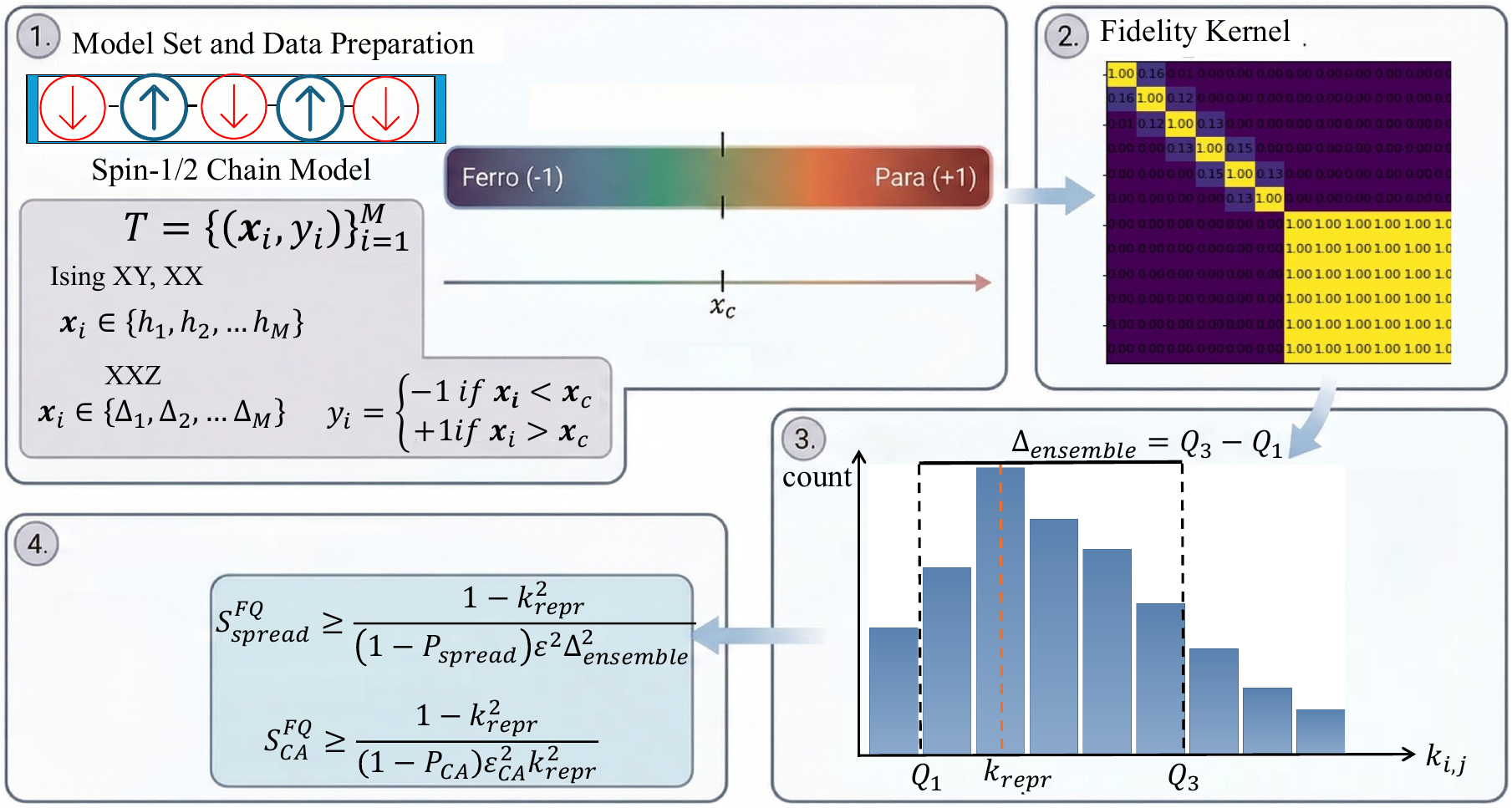}
    \caption{Schematic overview of the quantum kernel estimation pipeline for many-body spin chain systems.
   \textbf{(1)} \emph{Hamiltonian, phase labeling, and training set construction:} the anisotropic spin-chain Hamiltonian encompasses the Ising, XY, XX, XXZ models through the parameters $(\gamma ,\Delta, h)$. Training samples are chosen from two parameter windows on opposite sides of the critical point $\bm{x}_c$, assigning the labels $y_i = -1$ (ordered phase) and $y_i = +1$ (disordered phase). For the Ising universality class, the control parameter is $\bm{x}_i \equiv h_i$ while for XXZ it is $\bm{x}_i \equiv\Delta_i$.
   \textbf{(2)} \emph{Fidelity kernel and Gram matrix:} the labeled samples are used to build the fidelity-based kernel matrix (Gram matrix) with entries $k_{i,j}=|\langle \psi(\bm{x}_i)|\psi(\bm{x}_j)\rangle|^2$, where $| \psi(\bm{x}_j)\rangle$ is the ground state for model parameters $\bm{x}_j$.
    \textbf{(3)} \emph{Distribution of the elements of the Gram matrix:} 
    the information about the phase transition is encoded in the entries of the Gram matrix. To visualize the distribution of the entries we can use a histogram of their values between $0$ and $1$. The median value is denoted as $k_{\mathrm{repr}}$, and the spread of the distribution is quantified by the inter-quartile range (IQR)
    $\Delta_{\mathrm{ensemble}} = Q_3-Q_1$, where $Q_1$ and $Q_3$ are the first and third quartiles of the distribution.
    \textbf{(4)} \emph{Finite-size resource bounds:} each fidelity $k_{i,j}$ may be estimated experimentally with a SWAP test. 
    Given a finite number of experimental measurements, the estimated fidelity is subject to statistical fluctuations. The Chebyshev's inequality gives us a bound on the minimum number of measurements needed to ensure that the statistical fluctuations are small compared with the natural distribution of the Gram matrix elements. 
    }
    \label{fig:pipeline}
\end{figure*}

In this work we present a pathway to incorporate the quantum digital simulation of critical systems with a process that automatically 
achieves the classification of its phases via the computation of suitable metrics directly computable on a quantum processor. 
The operational pipeline is schematically depicted in Fig.~\ref{fig:pipeline}. 
In a digital simulation scenario, the ground state of a system can be faithfully prepared on a processor~\cite{VQE_lmg, Kiss2025}, 
on which observables, catching the critical behavior, are measured. 
Albeit it is a substantiated approach in model-targeted analog simulations, 
it falls short for digital ones and when little or no a priori information on the system under scrutiny is available. 
For such a reason, we apply an agnostic metric that relies only on geometric aspects of the quantum states. 
Namely, we present an analysis of SVMs with fidelity kernels and we benchmark it to 
the broader class of tunable spin-$\frac{1}{2}$ chain in a magnetic field. 
Our analysis delves into the feasibility of the learning protocol and its resource scaling, 
we underline the impact of the broken symmetry in the shot requirements and kernel concentration. 

Our objective is not to propose a universal algorithm for discovering previously unknown quantum phases. 
Instead, we focus on the practically relevant supervised-learning setting in which representative states from different phases are already available, 
for example, from quantum simulation, variational ground-state preparation, or experimentally calibrated parameter regimes. 
In this context, the central challenge is the quantum implementation of the learning protocol itself. 
Since the dominant computational cost arises from estimating the fidelity kernel, 
understanding how this cost scales with system size and with the symmetry of the underlying many-body system 
is essential for assessing the feasibility of fidelity-based quantum machine-learning algorithms.

\section{Theoretical framework}

In this section, we first introduce the Hamiltonian models for the critical quantum spin chains under investigation. Then we review the quantum kernel methods, based on fidelity and fidelity per site, that we are going to exploit for the classification of different phases and to identify the quantum phase transitions.

\subsection{Anisotropic spin-$1/2$ models and symmetry structure}
\label{sec:methods}

Since the early stages of quantum statistical mechanics, one-dimensional systems composed of interacting spins have served as paradigmatic models to study the emergence of complex structures in condensed matter. Due to their simple formulation and direct mapping of spin-$1/2$ degrees of freedom to qubits, they became pivotal testbeds for new techniques and associated resource costs, especially in quantum machine learning \cite{VQE_lmg, Barone_2024}. In the present work, we explore the suitability of fidelity-based support vector machines via a benchmark on spin-$1/2$ model with short range interaction,
defined by an anisotropic nearest-neighbor Hamiltonian 
\begin{align}
H = & -\sum_{i=1}^N \Big[
\frac{1+\gamma}{2}\, \sigma_i^x \sigma_{i+1}^x
+ \frac{1-\gamma}{2}\, \sigma_i^y \sigma_{i+1}^y
+ \Delta\, \sigma_i^z \sigma_{i+1}^z \Big] \nonumber\\
&- h \sum_{i=1}^{N} \sigma_i^z \, ,
\label{eq:hamiltonian}
\end{align}
having imposed PBC, such that $\sigma_{N+1} \equiv \sigma_1$ and with the Hamiltonian defined in units of a coupling constant $J$. This provides a unified description of several paradigmatic one-dimensional models
upon tuning the anisotropy parameters, $\gamma$ (for the $xy$ plane) and $\Delta$ (for the longitudinal $zz$ interaction), and the transverse field $h\ge 0$. 

For $\Delta=0$, one recovers the quantum XY chain, and the special points $\gamma=1$ and $\gamma=0$ correspond, respectively, to the quantum Ising chain and to the isotropic XX model. 
For $0<\gamma \leq1$, the model exhibits a continuous quantum phase transition in the Ising universality class between an ordered phase with nonvanishing magnetization along $x$ and a quantum paramagnet at $h > h_c = 1$, where $h_c$ is the critical field. In the ordered phase, the ground state is two-fold degenerate, reflecting the spontaneous breaking of a global spin-flip $\mathbb{Z}_2$ symmetry in the thermodynamic limit. However, for finite $N$ the ground state is typically unique, with an exponentially small splitting between the lowest parity sectors, with an enhanced splitting due to translation symmetry for PBC.
For $0<\gamma\leq 1$, the model has only a discrete $Z_2$ symmetry, while the continuous $U(1)$ symmetry is recovered only in the $\gamma=0$ XX limit, where it admits a field-driven quantum phase transition
separating a gapped paramagnetic phase in the region $h>1$, from a gapless critical phase along all the region $h \leq 1$ \cite{Franchini_2017}. 

For $\Delta\neq 0$ and $\gamma=0$, the Hamiltonian reduces to the XXZ Heisenberg chain. 
In our sign convention, the zero-field phase diagram displays a gapless phase for $- 1/2 < \Delta < 1/2$, separated by a Berezinskii--Kosterlitz--Thouless transition at $\Delta = - 1/2$ from a gapped antiferromagnetic regime, while $\Delta = 1/2$ marks a first-order transition to the ferromagnetic phase.

The field term $-h\sum_i \sigma_i^z$ tunes the competition between the exchange couplings and the polarization along the $z$ direction, and it is the control term used in the Ising/XY/XX analyses discussed below. Its symmetry effect depends on the anisotropy sector: for $\gamma=0$ (XX/XXZ), the Hamiltonian preserves the $U(1)$ symmetry associated with the conservation of $S^z_{\mathrm{tot}}$. 
In this work, we do not attempt a complete finite-field phase diagram of Eq.~\eqref{eq:hamiltonian}. We rather focus on the field-driven transitions in the Ising/XY/XX limits and on the zero-field XXZ transition structure relevant to our kernel analysis.

Throughout the numerical analysis, the XX and XXZ models (formally $\gamma=0$) are represented by a small nonzero anisotropy $\gamma=10^{-3}$, which regularizes the free-fermion expressions at the gap-closing momentum while leaving the $U(1)$/$SU(2)$ symmetry structure and the phase diagram unchanged. We use $\gamma=0$ when referring to the models definition and $\gamma=10^{-3}$ when quoting the parameters of a specific computation.

\label{subsec:phases}

\subsection{SVM with fidelity kernels}
\label{SVM_fid}

Given training data $\{(\bm{x}_i,y_i)\}_{i=1}^M$ with $\bm{x}_i\in\mathbb{R}^{n}$ and labels $y_i\in\{-1,+1\}$, an SVM learns a maximum-margin separating hyperplane. The hard-margin primal problem can be written using the Lagrangian
\begin{equation}
\mathcal{L}(\bm{w},b,\bm{\alpha})
= \tfrac{1}{2}\|\bm{w}\|^2 + \sum_{i=1}^M \alpha_i\bigl[1 - y_i(\bm{w}^\top \bm{x}_i + b)\bigr] \, ,
\label{eq:svm_lagrangian}
\end{equation}
where $\alpha_i$ are Lagrange multipliers, $\bm{w}\in\mathbb{R}^{n}$ is the weight vector, and $b$ is the bias. 
Stationarity with respect to $\bm{w}$ and $b$ implies $\bm{w}=\sum_{i=1}^M \alpha_i y_i \bm{x}_i$ and $\sum_{i=1}^M \alpha_i y_i=0$. 
Substituting these conditions into $\mathcal{L}$ yields a dual problem depending only on $\alpha$,
\begin{equation}
    \mathcal{L}_D(\alpha) = \sum_i \alpha_i - \tfrac{1}{2}\sum_{i,j} \alpha_i \alpha_j y_i y_j \, \bm{x}_i^\top \bm{x}_j \, ,
    \label{Eq2}
\end{equation}
and this objective function must be maximized over $\alpha_i \ge 0,$ subject to the constraint $\sum_i \alpha_i y_i = 0$. Typically, only a few $\alpha_i$ are non-zero, and the corresponding data points are known as support vectors. Therefore, the Lagrange multipliers $\alpha_i$ can be obtained by maximizing $\mathcal{L}_D(\alpha)$, and the support vectors allow for the estimation of the bias  via 
$b=\frac{1}{|\mathrm{SV}|}\sum_{k\in \mathrm{SV}}\left(y_k-\sum_{i\in \mathrm{SV}}\alpha_i y_i  \, \bm{x}_i^\top \bm{x}_k \right),$
where $\mathrm{SV}$ denotes the set of support vectors, such that 
$\mathrm{SV}=\left\{\, i \in \{1,\ldots,M\} \;:\; \alpha_i>0 \,\right\}$, and $|\mathrm{SV}|$ is the number of support vectors.
Although the above formulation describes a linear classifier, the true power of SVMs arises when the data are not linearly separable. In such cases, there may exist a map $\phi$ from the original input space to a higher-dimensional feature space, $\bm{x}_i \rightarrow \phi(\bm{x}_i)$, such that the data becomes linearly separable. 
A key advantage is that we do not need to compute this mapping explicitly, but we only need to evaluate inner products in the feature space, i.e.~$\langle \phi( \bm{x}_i ), \phi( \bm{x}_j )\rangle$, which can be computed through a kernel function $K( \bm{x}_i, \bm{x}_j)$. 
Replacing the inner product of the input space with a kernel function $K(\bm{x}_i,\bm{x}_j)$ yields
\begin{equation}
\mathcal{L}_D(\bm{\alpha})
= \sum_{i=1}^M \alpha_i
-\tfrac{1}{2}\sum_{i,j=1}^M \alpha_i\alpha_j y_i y_j\,K(\bm{x}_i,\bm{x}_j).
\label{eq:svm_dual_kernel}
\end{equation}
A kernel $K(\bm{x}_i,\bm{x}_j)$ quantifies similarity between inputs.  If $K$ is symmetric and positive semidefinite, Mercer’s theorem guarantees a feature map $\phi$ exists such that $K(\bm{x}_i,\bm{x}_j)=\langle \phi(\bm{x}_i),\phi(\bm{x}_j)\rangle$ \cite{scholkopf2002learning}. Thus, evaluating $K$ in input space is equivalent to computing an inner product in the (possibly high-dimensional) feature space.

\begin{figure}[t]
\centering
\begin{quantikz}[row sep=0.5cm, column sep=0.5cm]
\lstick{$\ket{0}$} & \gate{H} & \ctrl{1} & \gate{H} & \meter{} & \rstick{$\langle \sigma^z \rangle$} \\
\lstick{$\ket{\psi(h_i)}^{\otimes N}$} & \qwbundle{N} & \swap{1} & \qw & \qw & \qw \\
\lstick{$\ket{\psi(h_j)}^{\otimes N}$} & \qwbundle{N} & \targX{} & \qw & \qw & \qw
\end{quantikz}
\caption{Circuit diagram of the SWAP test employed to estimate 
the kernel entry $K(h_i, h_j) = |\braket{\psi(h_i)}{\psi(h_j)}|^2$. 
An ancilla qubit controls $N$ parallel SWAP gates, each acting on the corresponding pair of qubits from the two input registers. 
Measurement of the ancilla qubit yields the expectation value 
$\langle \sigma^z \rangle$, which is linearly related to the fidelity.}
\label{fig:swap-test-8q}
\end{figure}

In this work we employ a quantum-information-inspired similarity measure as the kernel. 
To this end, we introduce an inner product suited for the embedding of quantum many-body ground states. 
A natural choice is the Hilbert--Schmidt inner product,
$\langle A,B\rangle_{\mathrm{HS}}=\mathrm{Tr}(A^\dagger B)$. We take the feature map to be the rank-one projector $\phi(\bm{x})\equiv \rho(\bm{x})=\ket{\psi(\bm{x})}\bra{\psi(\bm{x})}$
on a $\bm{x}$-dependent wave function $\ket{\psi(\bm{x})}$, such that
\begin{equation}
K(\bm{x}_i,\bm{x}_j)=\langle \rho(\bm{x}_i),\rho(\bm{x}_j)\rangle_{\mathrm{HS}}
=\mathrm{Tr}\!\big[\rho(\bm{x}_i)\rho(\bm{x}_j)\big].
\label{eq:hs_kernel}
\end{equation}
For pure states, this kernel reduces to the fidelity:
\begin{equation}
K(\bm{x}_i,\bm{x}_j) := \left| \braket{\psi(\bm{x}_i)}{\psi(\bm{x}_j)} \right|^2.
\label{eq:global}
\end{equation}

By construction, the fidelity kernel is positive semidefinite and yields a symmetric positive-semidefinite (PSD) Gram matrix for any dataset. 
In the following, the model parameters serve as the data points, i.e.~$\bm{x} = (h, \Delta, \gamma )$,
that define the quantum feature map through the corresponding ground state $\ket{\psi(\bm{x})}$.
The \textit{signed decision} function is given by
\begin{equation}
    d(\bm{x})=\sum_{i=1}^{M} \alpha_i y_i\, K(\bm{x}_i,\bm{x})+b \, ,
    \label{eq:DF}
\end{equation}
where the sum is over the training data points.
Here $d(\bm{x})$ is termed signed because it indicates the predicted class. 
If $d( \bm{x} ) > 0$ the classifier predicts $\hat y( \bm{x} )=+1$, whereas if $d( \bm{x} ) < 0$ it predicts $\hat y( \bm{x} ) =-1$, i.e., 
$\hat y( \bm{x} )=\mathrm{sign}\!\left(d( \bm{x} )\right)$. 
The decision boundary is defined by the hyperplane $d(\bm{x})=0$. 
Furthermore, the magnitude $|d(\bm{x})|$ proportional to the distance from the hyperplane in feature space, serving as a measure of classification confidence.

We also consider an alternative kernel defined from the fidelity per site:
\begin{equation}\label{eq:fid_per_site}
f_N(\bm{x}_i,\bm{x}_j) = \big[F(\bm{x}_i,\bm{x}_j)\big]^{1/N} \, .
\end{equation}
Using the fidelity per site is sometimes a better choice as the fidelity may tend to decrease exponentially with increasing system size $N$ \cite{Gu2010Review}.
This fidelity per site is also helpful when comparing different system sizes on the same scale.

\begin{figure}[t]
\centering
\includegraphics[width=\linewidth]{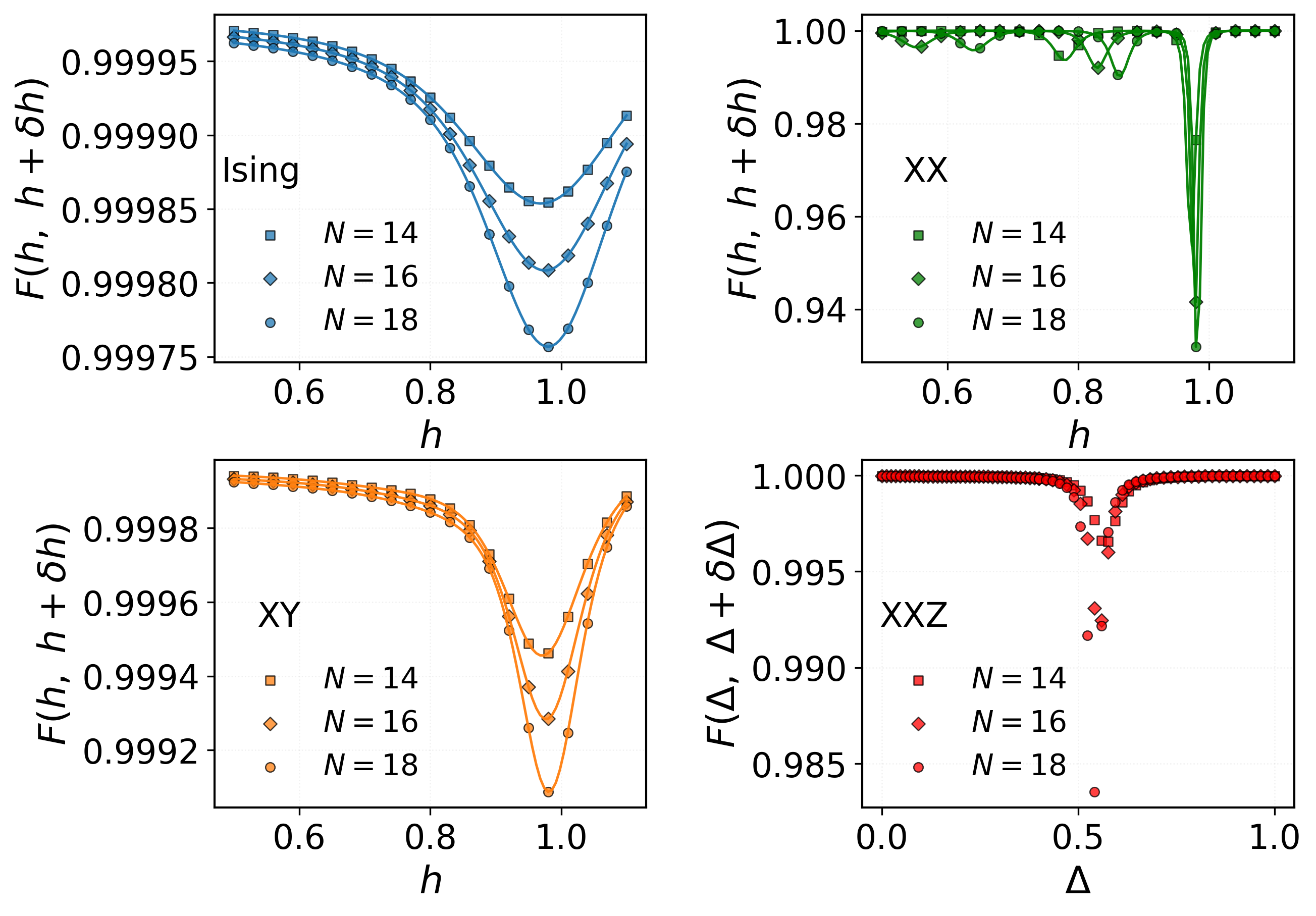}
\caption{Fidelity $F(x,x{+}\delta x)$ vs control parameter for $N=14,16,18$.
The dip marks the transition:  near $h\!\approx\!1$ for  (Ising/XX/XY),
 and  near $\Delta\!\approx\!0.5$ for (XXZ, $h=0$, first-order boundary).
The BKT boundary at $\Delta\!\approx\!-0.5$ is analyzed in Sec.~\ref{sec:xxz_bkt}.}
\label{fig:critical_points}
\end{figure}

\begin{figure*}
    \centering
    \includegraphics[width=0.99\textwidth]{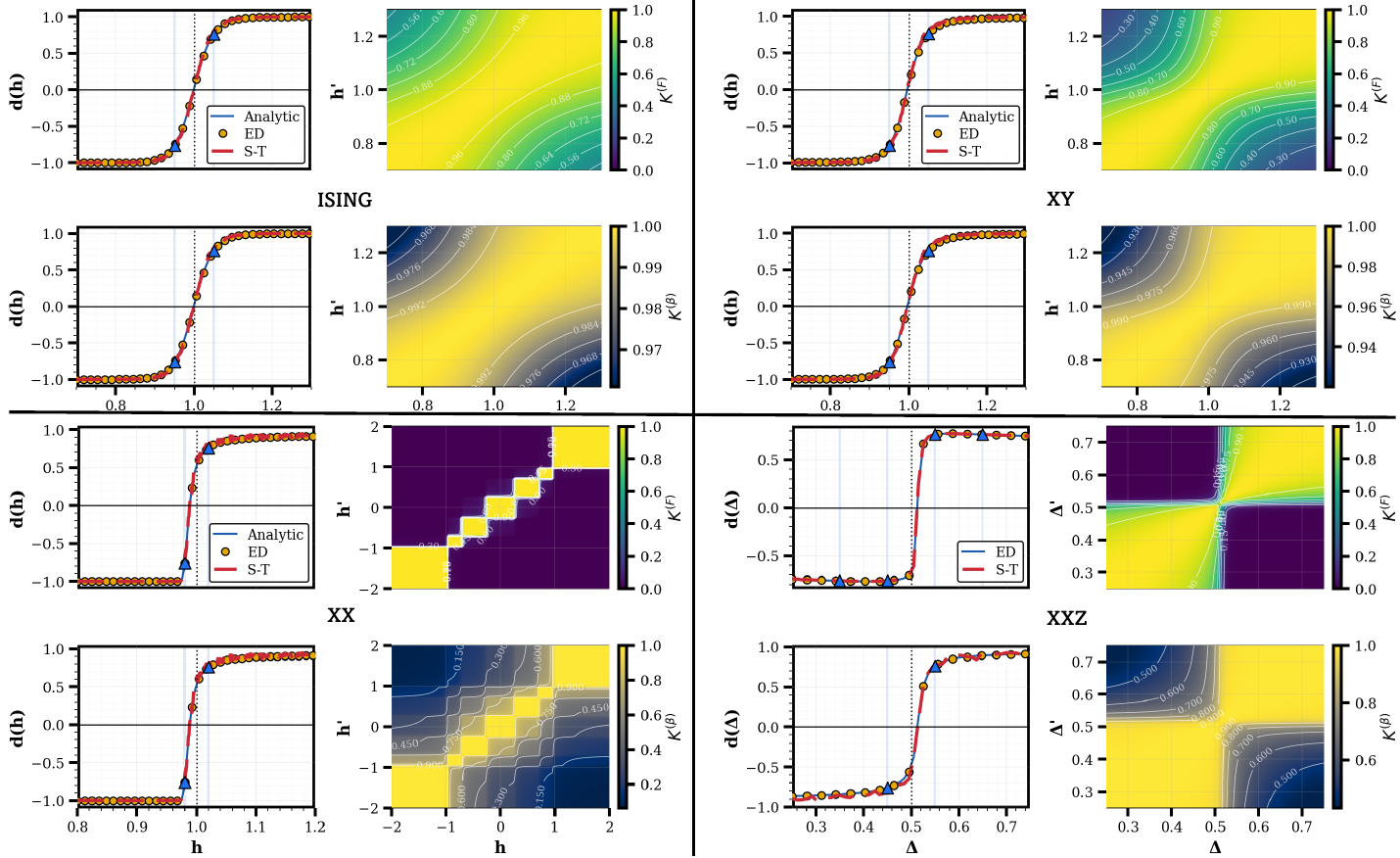}
    \caption{The figure shows the signed SVM decision functions and the corresponding kernel matrices, for four Hamiltonian models.
    The signed SVM decision functions, introduced in Eq.~\eqref{eq:DF}, are shown in the line plots in the first and third columns. The kernel matrices are shown in the heat map plots in the second and fourth columns.
    In the decision-function panels, the horizontal axis is the control parameter ($h$ for Ising/XY/XX, and $\Delta$ for XXZ), and the vertical axis is the signed decision function value $d(x)$. 
    In the kernel-matrix panels, the horizontal and vertical axes label the pairwise control parameters used to build the Gram matrix ($h,h'$ for Ising/XY/XX, and
    $\Delta,\Delta'$ for XXZ), and the color scale gives the kernel value. 
    SWAP-test (ST) estimates are benchmarked against numerical exact diagonalization (ED), and against the analytical result, when available (i.e.~for the XY/XX cases). 
    The vertical dotted line marks the expected critical point. %
    For each model, the upper panels (first and third rows) show the kernel matrix defined from the global fidelity kernel, while the lower panels correspond to fidelity per site (see Eq.~\eqref{eq:fid_per_site}).
    For the Ising/XY/XX panels, the SVM is trained on two field windows around the critical point $h_c=1$, namely $h \in [0.7,0.95] \cup [1.05,1.3]$, using 16 training points per side ($M=32$ total). 
    For the XXZ panels (at $h=0$), the SVM is trained on two anisotropy windows around the critical point 
    $\Delta_c=0.5$, namely 
    $\Delta\in[0.35,0.45]\cup[0.55,0.65]$, using 16 training points per side ($M=32$ total). 
    For the XX case, we display the kernel matrix over a broad field range that includes both the gapless region ($h<1$) and the saturated phase ($h>1$), and we choose the training set from the same two windows around $h_c=1$ as in the Ising/XY cases.}
\label{fig:svm_combined}
\end{figure*}

\subsection{SVM for quantum phase classification}
\label{SVM_class}

The phase identification task can be viewed as a classification problem. In any realistic scenario involving systems with different phases, the target is to determine the critical points in the parameter space where the phase transitions occur. 
Due to finite resources, we may only estimate these points for finite system size $N$. We denote these finite-size pseudo-critical points as $h_c(N)$ and $\Delta_c(N)$, respectively. These pseudo-critical estimates are used to set the training windows on the two sides of the transition when constructing the labeled dataset (see Fig.~\ref{fig:pipeline}). 
A well-known way to locate the phase transition is through the fidelity of nearby ground states~\cite{ZanardiPaunkovic2006, CamposVenutiZan07, Zanardi_2007}, separated by a small parameter step $\delta \bm{x}$.
We denote such a fidelity as 
$F(\bm{x},\bm{x}+\delta \bm{x})\equiv K(\bm{x},\bm{x}+\delta \bm{x})$,
where $\bm{x}$ is the control parameter (either $h$ or $\Delta$ in our settings). At the phase transition, such fidelity is expected to show a singular behavior.
Figure~\ref{fig:critical_points} shows the nearest-neighbor fidelity $F(\bm{x},\bm{x}+\delta \bm{x})$ as a function of the control parameter $\bm{x}$ for several models and system sizes. In the Ising/XY/XX cases (panels a--c) we take $\bm{x}=h$, while for the XXZ chain at $h=0$ (panel d) we take $\bm{x}=\Delta$.
In all cases the transition region is signaled by a clear minimum of $F(\bm{x},\bm{x}+\delta \bm{x})$,
meaning that ground states separated by an infinitesimal parameter step become most distinguishable there.

The main objective of our work is to characterize SVM based on fidelity for the classification of quantum phase transitions. This implies estimating the kernel matrix, whose entries are the fidelities between ground states corresponding to different points in the training set. 
We emphasize that we chose such a quantum-information-informed kernel because it does not require any prior assumptions about order parameters or correlation functions. It offers a significant advantage for characterizing quantum phase transitions, but it is well known that its scalability can be challenging on actual devices. 
In  case of the paradigmatic ferro/para-magnetic transition, the signed decision function results in a powerful tool to quantitatively locate it~\cite{Giannetti_2019}. Here we are extending this approach to gapless models such as XX and XXZ.
For finite $N$, the position of this minimum defines a pseudo-critical estimate, whose shift with $N$ captures finite-size corrections.
Therefore, a rigorous analysis of resource scaling for systems of finite size is compelling to benchmark the usefulness of state-of-the-art quantum processors~\cite{Kiss2025}. 

Before proceeding to the finite-size scaling analysis of the SVM estimates, it is useful to visualize the full output of the fidelity kernel framework for all models. 
Figure~\ref{fig:svm_combined} reports the signed decision function $d(\bm{x})$ in Eq.~\eqref{eq:DF} (left panels) and the corresponding Gram matrices (contour plots in the right panels). 
In the left panels we also compare different methods to compute the kernel matrix: SWAP-test estimates, exact numerical diagonalization of the Hamiltonian, and, when available, the analytical form of the ground states. 
The vertical dotted line marks the expected critical points, where the decision function changes its sign, and the classifier switches phase labels. 
For each Hamiltonian model, the upper-right panels show the global fidelity kernel, whereas the lower-right panels show the fidelity-per-site kernel, introduced in Eq.~\eqref{eq:fid_per_site}. 
Notably, the decision function exhibits sharp behavior in the XX model for $h \leq 1.$ Unlike other models with isolated critical points, the entire line $(\gamma=0, h\leq 1)$ is critical here. 
This manifests as a prominent central high-fidelity band displaying a distinctive staircase-like pattern.

\section{Finite-size scaling of the computational resources}

In the previous section, we have established that fidelity kernels accurately identify the phase boundaries. 
We now address the main objective of this work: the resource requirements associated with implementing this learning protocol on quantum hardware. 
In contrast to previous studies \cite{SanchoLorente2022PRA}, our emphasis is not on improving the classifier itself 
but on quantifying the measurement cost required to estimate the fidelity kernel under realistic finite-shot conditions.

The training of the SVM requires first to collect a training dataset of size $M$, and then to estimate the elements of the corresponding $M \times M$ Gram matrix, which, we recall is defined from the fidelity between different ground states.
In an experimental setup, the fidelity is estimated using the SWAP test, whose circuit diagram is shown in Fig.~\ref{fig:swap-test-8q}. In the SWAP test, the probability $P_0$ of obtaining the outcome $|0\rangle$ when measuring the control qubit is related to the fidelity as
\begin{equation}
\label{eq:K_swap}
    2P_0 - 1 = K(\bm{x}_i,\bm{x}_j) \, .
\end{equation}
However, for any finite number of runs of the circuit, one can only obtain the empirical relative frequency of outcome $|0\rangle$, from which the probability $P_0$ can only be estimated with a corresponding statistical error due to finite-size fluctuations.

In this section, we explore how the SVM performances depend on the choice and the size $M$ of the training set, and on the number of repeated SWAP tests needed to estimate the fidelity. Both these quantities will depend on the size $N$ of the given spin chain, and crucially on the symmetry of the considered Hamiltonian model.

\subsection{Size and properties of the training set}
\label{SVM_train}

\begin{figure}[t]
    \centering
    \includegraphics[width=0.9\linewidth]{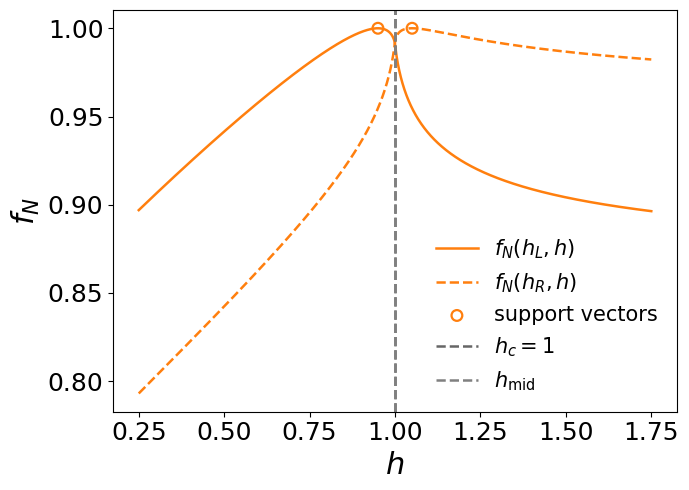}
    \caption{Decision boundary as a geometric midpoint for the per-site fidelity kernel (XY chain, $\gamma=0.5$).
We train an SVM with a precomputed per-site kernel $f_N(h,h')=F(h,h')^{1/N}$ on two field windows
$h\in[0.76,0.95]\cup[1.05,1.30]$ (16 points per side, $M=32$ total), using the analytical Bogoliubov-angle fidelity at $N=1000$.
The solid and dashed curves show the similarities to the \emph{inner} (dominant) support vectors, plotted as $f_N(h_L,h)$ and $f_N(h_R,h)$, where
$h_L=\max\{h_i:\alpha_i>0,y_i=-1\}$ and $h_R=\min\{h_i:\alpha_i>0,y_i=+1\}$.
Open circles mark the two dominant support vectors.
The dashed vertical line marks the expected critical field $h_c=1$, while the dotted line marks the midpoint estimate
$h_{\mathrm{mid}}$ defined by $f_N(h_L,h_{\mathrm{mid}})=f_N(h_R,h_{\mathrm{mid}})$.}
    \label{fig:Boundary}

\end{figure}
\begin{figure}[h]
    \centering
    \includegraphics[width=0.9\linewidth]{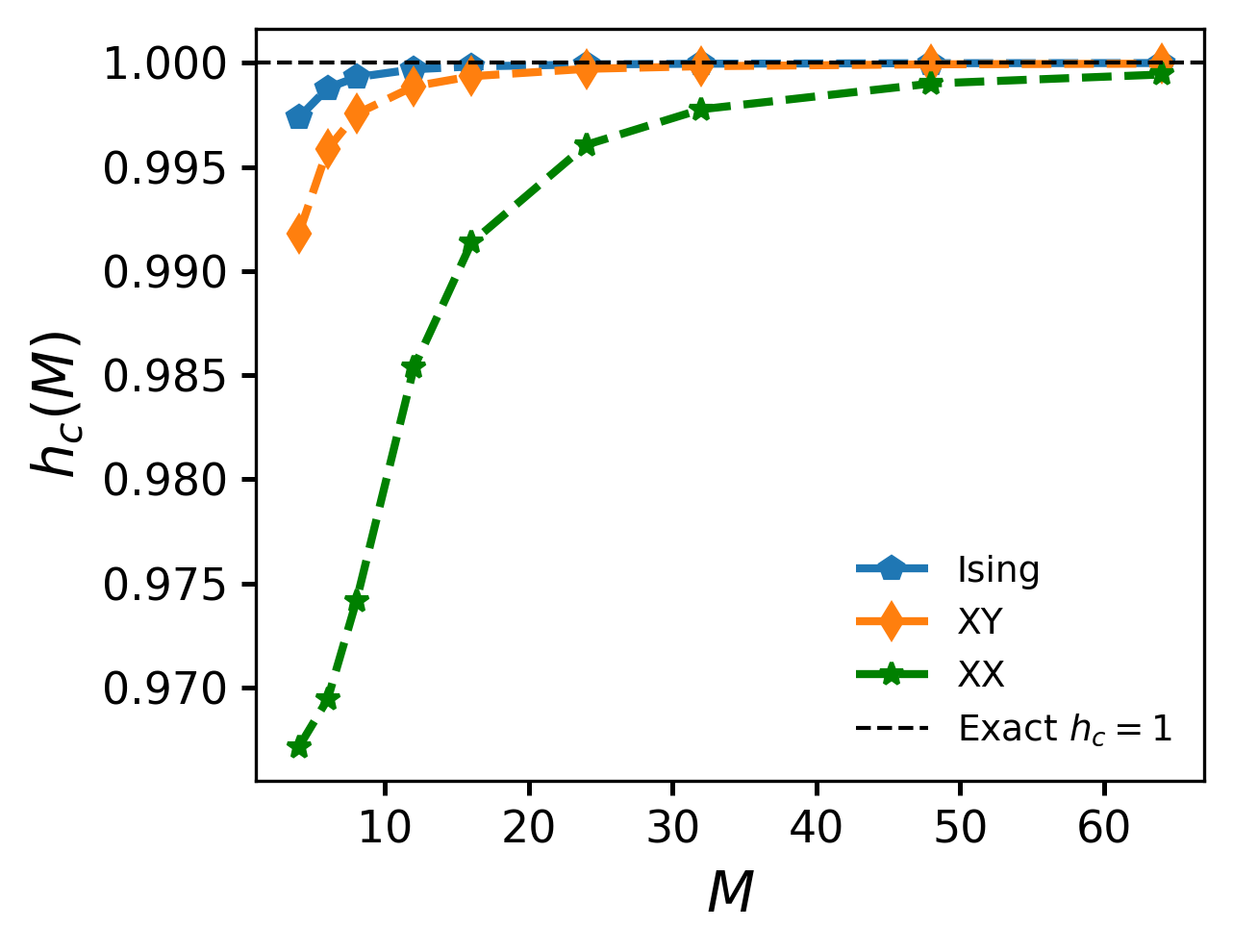}
    \includegraphics[width=0.9\linewidth]{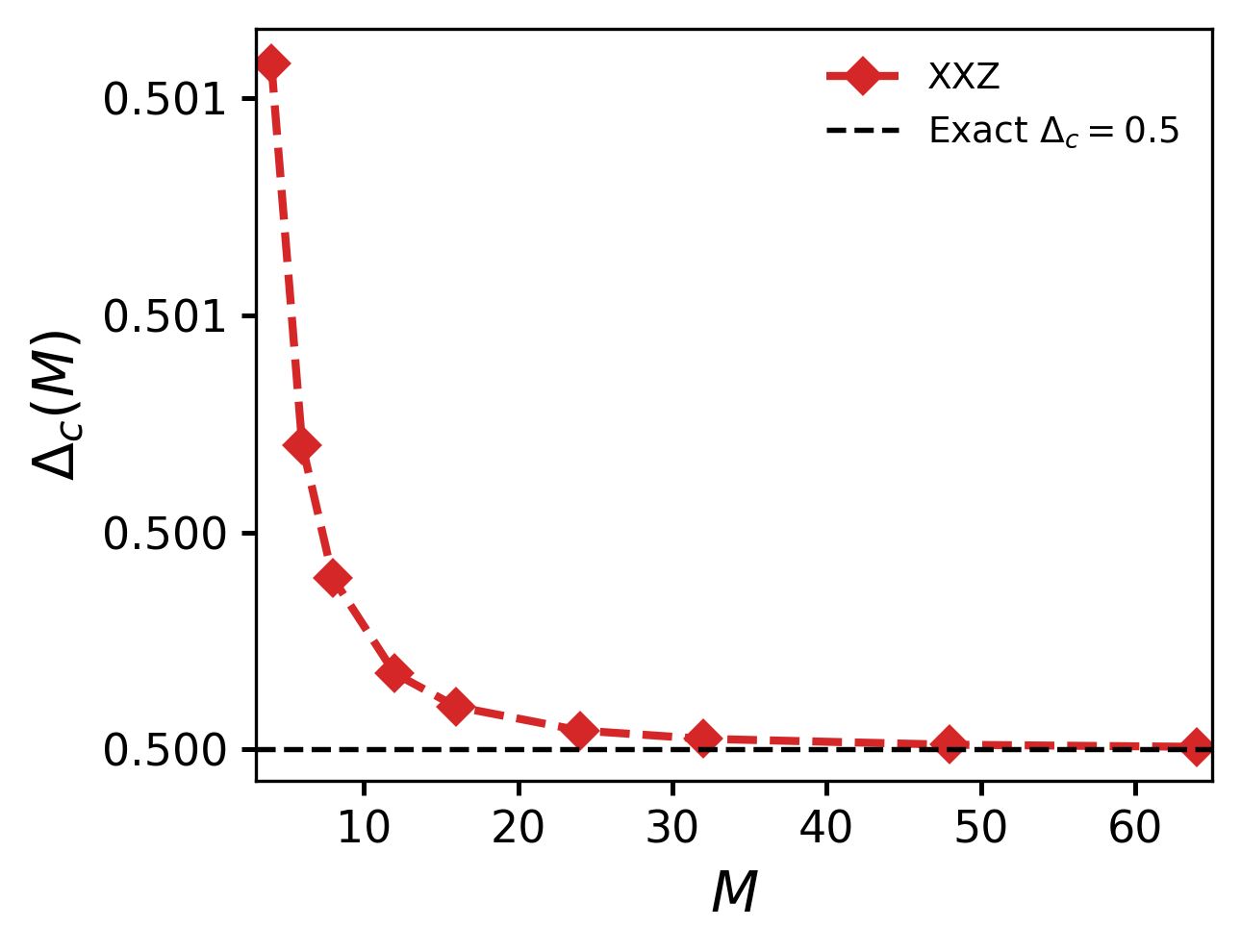}
    \caption{SVM estimates of the critical point versus the  training-set size $M$.
    The upper panel is for the Ising model ($\gamma=1$, $\Delta=0$), the XY model ($\gamma=0.5$, $\Delta=0$), and the XX model ($\gamma=10^{-3}$, $\Delta=0$), for $N=12$ sites. For these models we have chosen $M$ training points equally spaced within the interval $h \in [0.85, 1.15]$, which is centered around the transition point $h_c=1$.
    The lower panel is for the XXZ model ($\gamma=10^{-3}$, $h=0$). For this model we have chosen $M$ training points equally spaced within the interval $\Delta \in [0.35, 0.65]$, which is centered around the transition point $\Delta_c = 0.5$.
    The black dashed lines mark the exact critical values ($h_c=1$ in the upper panel and $\Delta_c=0.5$ in the lower panel).}
    \label{fig:h_c_vs_M}
\end{figure}

In general, one would expect that increasing the size $M$ of the training dataset will lead to improved accuracy in the determination of the critical point. Moreover, the way in which the training points are selected could affect the accuracy, for given size to the dataset.

We have investigated the relation between accuracy and the properties and size of the training dataset, for the four Hamiltonian models under our investigation. We have considered equally spaced training data points chosen within an interval centered around the transition point. As a general trend, as  it should expected, we find that increasing the number of $M$ training points leads to higher accuracy. This is shown in Fig.~\ref{fig:h_c_vs_M}. 
A similar trend is obtained using randomly chosen training data points within the same interval. In both cases, we observe that models with increasing symmetry (we recall that the symmetry increases when moving from Ising to XY and to XX models) require a larger training data set to achieve comparable accuracy in the determination of the estimated transition point. 

To achieve a better understanding of the relation between accuracy and the size $M$ of the training dataset, we need to investigate where the SVM places the separating hyperplane. This is related to the support vectors. Here we focus on the \emph{dominant} support vectors, which are defined as those support vectors that mainly determine the boundary: i.e.~either they correspond to points closest to the critical region in parameter space, or those with the largest $\alpha_i$ (or both).
Then we consider the case where there are two dominant vectors: one from the left interval and one from the right interval around the true critical point $\bm{x}_c$, denoted as $\bm{x}_L$ and $\bm{x}_R$, respectively. 
Also, we denote as $\alpha_L$ and $\alpha_R$ the corresponding nonzero dual coefficients, and the corresponding ground states are $\ket{\psi(\bm{x}_L)}$ and $\ket{\psi(\bm{x}_R)}$. 
Their mutual distance is~\cite{scholkopf2000kernel}
\begin{align}
  D(\bm{x}_L, \bm{x}_R) &= \|\phi(\bm{x}_L) - \phi(\bm{x}_R)\| \nonumber \\
    &= \sqrt{K(\bm{x}_L,\bm{x}_L)+K(\bm{x}_R,\bm{x}_R)-2K(\bm{x}_L, \bm{x}_R)} \nonumber \\
    &= \sqrt{2(1 - K(\bm{x}_L, \bm{x}_R))} \, ,
\end{align}
where the last step uses the normalization $K(\bm{x},\bm{x})=1$ induced by the fidelity.
To determine the exact location of the decision boundary, we recall the dual constraint of the SVM optimization $\sum_i \alpha_i y_i = 0$. In our symmetric scenario with two dominant support vectors with labels, $y_L=-1,\, y_R=+1$, this constraint implies $\alpha_L = \alpha_R$. Consequently, in this symmetric idealized case the bias term vanishes ($b\simeq 0$), and the separating hyperplane is placed exactly at the geometric midpoint where the fidelities to the support vectors are balanced, namely $D(\bm{x}_L, \bm{x})= D(\bm{x}_R, \bm{x})$. In Fig.~\ref{fig:Boundary} the two curves show, for the XY model, the kernel similarities to dominant support vectors, plotted as $f_N(h_L,h)$ and $f_N(h_R,h)$.
The decision boundary corresponds to the geometric midpoint between $\phi(h_L)$ and $\phi(h_R)$, i.e., the point $h_{\mathrm{mid}}$ where the similarities balance,
$f_N(h_L,h_{\mathrm{mid}})=f_N(h_R,h_{\mathrm{mid}})$ (vertical dotted line), which is equivalent to equal distances
$\|\phi(h_L)-\phi(h_{\mathrm{mid}})\|=\|\phi(h_R)-\phi(h_{\mathrm{mid}})\|$.

As a consequence of that, training near the critical value yields a better estimation of the zero-crossing of the decision function (discussed in Eq.~\eqref{eq:DF}). As shown in Fig.~\ref{fig:h_c_vs_M}, for a fixed training interval, the SVM estimate of the critical point starts to settle as we increase the number of training points $M$. This is physically intuitive: when we have very few points, the SVM does not have enough information to capture what happens close to the transition, so the estimated boundary can shift. As we add more points, the SVM eventually identifies the most important ones, i.e., the support vectors. 

What matters in our construction is that the training set is sampled from two fixed parameter windows placed on opposite sides of the transition. Denote these windows as
$I_L=[\bm{x}_L^{\min},\bm{x}_L^{\max}]$ and $I_R=[\bm{x}_R^{\min},\bm{x}_R^{\max}]$, with $\bm{x}_L^{\max}<\bm{x}_R^{\min}$.
The \emph{endpoints} are therefore the four boundary values $\bm{x}_L^{\min},\bm{x}_L^{\max},\bm{x}_R^{\min},\bm{x}_R^{\max}$, which are known once the windows are chosen. The two values closest to the critical region are the \emph{inner endpoints}, $\bm{x}_L^{\max}$ and $\bm{x}_R^{\min}$, and they \emph{bracket} the transition in parameter space in the sense that the true critical point satisfies $\bm{x}_L^{\max}<\bm{x}_c<\bm{x}_R^{\min}$. Importantly, we do not know the support vectors a priori. The SVM identifies them during training as the samples with nonzero dual coefficients $\alpha_i>0$ that constrain the maximum--margin solution. In this one-dimensional and approximately symmetric setup, the inner endpoints $(\bm{x}_L^{\max},\bm{x}_R^{\min})$ are often selected among the dominant support vectors because they provide the tightest constraints near the transition. Once these constraints are present, adding more samples strictly inside the same intervals typically yields $\alpha_i=0$ for the additional points and does not change the separating hyperplane, so the estimated boundary saturates. Saturation tends to occur faster for models with discrete $\mathbb{Z}_2$ symmetry (Ising and XY) than for those with continuous $U(1)$ symmetry (XX and  XXZ).

\subsection{Sample complexity for the estimation of the Gram matrix}

Here we focus on assessing the relation between the accuracy in the estimation of the kernel and the number of repeated SWAP tests (referred to as measurement shots).

Recently, Miroszewski~\emph{et al.} have provided a bound to estimate the minimum number of shots needed to reliably resolve the spread of kernel values in a quantum kernel setting~\cite{miroszewski2024search}. The present work extends this framework in three directions. First, we adapt the shot bounds from a ZZ feature-map kernel to fidelity-based kernels estimated via the SWAP test. Second, through a closed-form evaluation of the mode overlap and of the Bogoliubov-angle sensitivity, as reported in Eqs.~\eqref{eq:mode_overlap_exact} and \eqref{eq:theta_derivative}, we establish a direct quantitative connection between the symmetry class of the model and the degree of kernel concentration, 
explaining the shot-cost hierarchy Ising~$<$~XY~$<$~XX from first principles. Third, we extend the analysis to the genuinely interacting XXZ-model and demonstrate that the enhanced $SU(2)$ symmetry near the isotropic point\footnote{At the isotropic point $\Delta=\frac12$ the model matches with the Heisenberg model that possesses SU(2) symmetry.}
further amplifies the resource cost beyond the free-fermion $U(1)$ case.

\subsubsection{Statistical model of the fidelity kernel estimator}

We recall from Sec.~\ref{SVM_fid} that the SVM receives as its sole
input the $M \times M$ Gram matrix, whose $(i,j)$ entry is the
fidelity kernel value $k_{i,j} = K(\bm{x}_i, \bm{x}_j)$ defined in
Eq.~\eqref{eq:global}. A point that deserves emphasis is that this
matrix is \emph{never} estimated as a whole on quantum hardware.
Each entry $k_{i,j}$ must be estimated independently by running the
SWAP-test circuit for the specific pair $(\bm{x}_i, \bm{x}_j)$
a finite number of times $S$. In the $\ell$-th shot, the ancilla
qubit yields a binary outcome $z_\ell \in \{0,1\}$, from which the
empirical estimator

\begin{equation}
\hat{P}_0 = \frac{1}{S}\sum_{\ell=1}^{S} \mathbf{1}_{\{z_\ell = 0\}}\end{equation}
is constructed, where $\mathbf{1}_{\{z_\ell=0\}}$ denotes the indicator function, equal to $1$ if the $\ell$-th shot returns $z_\ell=0$ and to $0$ otherwise, so that $\hat{P}_0$ is the relative frequency of the outcome $\ket{0}$ over the $S$ shots. The kernel estimate then follows via Eq.~\eqref{eq:K_swap}. Since $z_\ell$ is a Bernoulli random variable
with success probability $P_0 = (1+k_{i,j})/2$, the estimator
$\hat{k}_{i,j} = 2\hat{P}_0 - 1$ is unbiased and satisfies
\begin{equation}
\mathrm{Var}\!\left(\hat{k}_{i,j}\right) = \frac{1 - k_{i,j}^2}{S} \, .
\label{eq:swap_var}
\end{equation}
The shots  $S$ thus controls the statistical precision of each
entry separately, and filling the full Gram matrix requires
$O(M^2 S)$ SWAP-test runs in total. The diagonal entries
$k_{i,i} = 1$ are exact by normalization and require no estimation
all resource analysis below therefore concerns only the
$\binom{M}{2}$ off-diagonal entries.

We recall from Sec.~\ref{SVM_fid} that the signed decision function
$d(\bm{x})$ of Eq.~\eqref{eq:DF} is built entirely from pairwise
kernel evaluations. The Gram matrix encodes the geometry of the
training set in the kernel feature space, two parameter points
$\bm{x}_i$ and $\bm{x}_j$ from opposite phases will typically have a
small overlap $k_{i,j} <1$, while points within the same phase
will have $k_{i,j}$ closer to unity. Both the support vectors and
the bias $b$ are determined by this geometry alone. Finite-shot
noise therefore does not merely add a uniform uncertainty to the
predictions it distorts the \emph{relative} structure of the
matrix, potentially misidentifying which training points are close
or far in feature space, and thereby shifting the decision boundary.
The central question is therefore not whether any single entry can
be estimated with some accuracy in isolation, but whether the
\emph{contrast} of the Gram matrix the systematic variation of
kernel values across the dataset survives the statistical
fluctuations introduced by a finite number of shots.

\subsubsection{Kernel  statistics: representative scale and
ensemble spread}

To translate the above requirement into a concrete shot bound, we
need two summary statistics that characterize the off-diagonal
fidelity distribution: a \emph{location scale}, giving the typical
magnitude of the off-diagonal entries, and a \emph{spread scale},
quantifying the contrast available across the dataset.

Checking the pairwise separation $|k_{i,j} - k_{p,q}|$ for all
possible entry pairs in an $M \times M$ Gram matrix is clearly
impractical. Following Ref.~\cite{miroszewski2024search}, we
summarize the spread of the off-diagonal distribution by the
interquartile range (IQR),
\begin{equation}
\Delta_{\mathrm{ensemble}} = Q_3 - Q_1 \, ,
\end{equation}
where $Q_1$ and $Q_3$ are the first and third quartiles of
$\{k_{i,j} : i < j\}$. The IQR captures the range spanned by the
central $50\%$ of the off-diagonal entries and is less sensitive to
outliers than the full range, making it a robust measure of the
fidelity contrast available to the SVM.

For the location scale, we use the \emph{median} of the same
off-diagonal entries,
\begin{equation}
k_{\mathrm{repr}} = \mathrm{median}\!\left(\{k_{i,j} : i < j\}\right).
\label{eq:krepr}
\end{equation}
The choice of the median rather than the mean is deliberate and
physically motivated. The distribution of off-diagonal fidelities
for a many-body system is generically skewed: a small number of
same-phase pairs near a common reference point contribute entries
close to unity, while the bulk of cross-phase pairs accumulates near
zero as $N$ grows. In this regime the mean is pulled toward the
high-fidelity tail and no longer represents the entry that a generic
SWAP-test circuit will encounter. The median, being the value
exceeded by exactly half the off-diagonal entries, is robust to
these outliers and provides a conservative central benchmark for
shot counting. It also pairs naturally with the IQR, since both are
quartile-based statistics and therefore characterize the same central
bulk of the distribution. In our noiseless benchmark,
$k_{\mathrm{repr}}$ is obtained from exact off-diagonal fidelities
computed via exact diagonalization (ED) or from the analytical
Bogoliubov expressions of Appendix~\ref{app:bogoliubov_fidelity},
and $\Delta_{\mathrm{ensemble}}$ is estimated from the corresponding
quartiles. These two statistics lead to two physically distinct shot
bounds, which we now derive in turn.

\subsubsection*{Spread-resolution bound}

The first requirement is that the finite-shot noise on each entry
must not wash out the contrast between entries. We recall from
Eq.~\eqref{eq:swap_var} that the SWAP-test estimator is unbiased,
i.e.\ its mean value coincides with the true kernel entry,
$\mathbb{E}[\hat{k}_{i,j}]=k_{i,j}$. If the absolute deviation of
the estimator from this mean, $|\hat{k}_{i,j} - \mathbb{E}[\hat{k}_{i,j}]|$,
becomes comparable to $\Delta_{\mathrm{ensemble}}$, then entries that
genuinely differ by a typical amount become statistically
indistinguishable, the Gram matrix appears effectively uniform, and
the SVM loses its ability to place a meaningful decision boundary.
We therefore demand that
\begin{equation}
\Pr\!\left(|\hat{k}_{i,j} - \mathbb{E}[\hat{k}_{i,j}]| \ge
\varepsilon\,\Delta_{\mathrm{ensemble}}\right) \le 1 - P_{\mathrm{spread}} \, ,
\label{eq:spread_req}
\end{equation}
i.e.\ the deviation exceeds a fraction $\varepsilon$ of the IQR with
probability at most $1 - P_{\mathrm{spread}}$.

To turn Eq.~\eqref{eq:spread_req} into a concrete shot bound, we
recall Chebyshev's inequality: for any random variable $X$ with mean
$\mathbb{E}[X]$ and finite variance $\mathrm{Var}(X)$, and for any
threshold $t>0$,
\begin{equation}
\Pr\!\left(|X-\mathbb{E}[X]|\ge t\right)\le\frac{\mathrm{Var}(X)}{t^2} \, .
\label{eq:chebyshev}
\end{equation}
Applying Eq.~\eqref{eq:chebyshev} to the SWAP-test estimator
$X=\hat{k}_{i,j}$ with threshold
$t=\varepsilon\,\Delta_{\mathrm{ensemble}}$, and using the variance
from Eq.~\eqref{eq:swap_var}, gives
\begin{equation}
\Pr\!\left(|\hat{k}_{i,j} - \mathbb{E}[\hat{k}_{i,j}]| \ge
\varepsilon\,\Delta_{\mathrm{ensemble}}\right)
\le
\frac{1-k_{i,j}^2}{S\,\varepsilon^2\,\Delta_{\mathrm{ensemble}}^2} \, .
\label{eq:chebyshev_spread}
\end{equation}
Comparing Eq.~\eqref{eq:chebyshev_spread} with the requirement in
Eq.~\eqref{eq:spread_req}, it is sufficient to impose
\begin{equation}
\frac{1-k_{i,j}^2}{S\,\varepsilon^2\,\Delta_{\mathrm{ensemble}}^2}
\le 1-P_{\mathrm{spread}} \, ,
\end{equation}
which, solved for $S$, gives
\begin{equation}
S\ge\frac{1-k_{i,j}^2}{(1-P_{\mathrm{spread}})\,\varepsilon^2\,
\Delta_{\mathrm{ensemble}}^2} \, .
\end{equation}
Finally, substituting $k_{i,j}$ by the representative value
$k_{\mathrm{repr}}$ of Eq.~\eqref{eq:krepr} yields the
spread-resolution shot bound
\begin{equation}
S_{\mathrm{spread}} \ge
\frac{1 - k_{\mathrm{repr}}^2}
{(1 - P_{\mathrm{spread}})\,\varepsilon^2\,
\Delta_{\mathrm{ensemble}}^2} \, .
\label{eq:Sspread-main}
\end{equation}
The numerator $1 - k_{\mathrm{repr}}^2$ sets the intrinsic noise
level of the SWAP-test estimator at the representative fidelity
scale. In the denominator, a stricter tolerance $\varepsilon$
increases the cost as $1/\varepsilon^2$, while kernel
concentration smaller $\Delta_{\mathrm{ensemble}}$ increases it
as $1/\Delta_{\mathrm{ensemble}}^2$. This second dependence is the
central diagnostic for the symmetry-class comparison carried out in
the  section \ref{results}.

\subsubsection*{Concentration-avoidance bound}

\begin{figure*}
    \centering
    \includegraphics[width=0.85\linewidth]{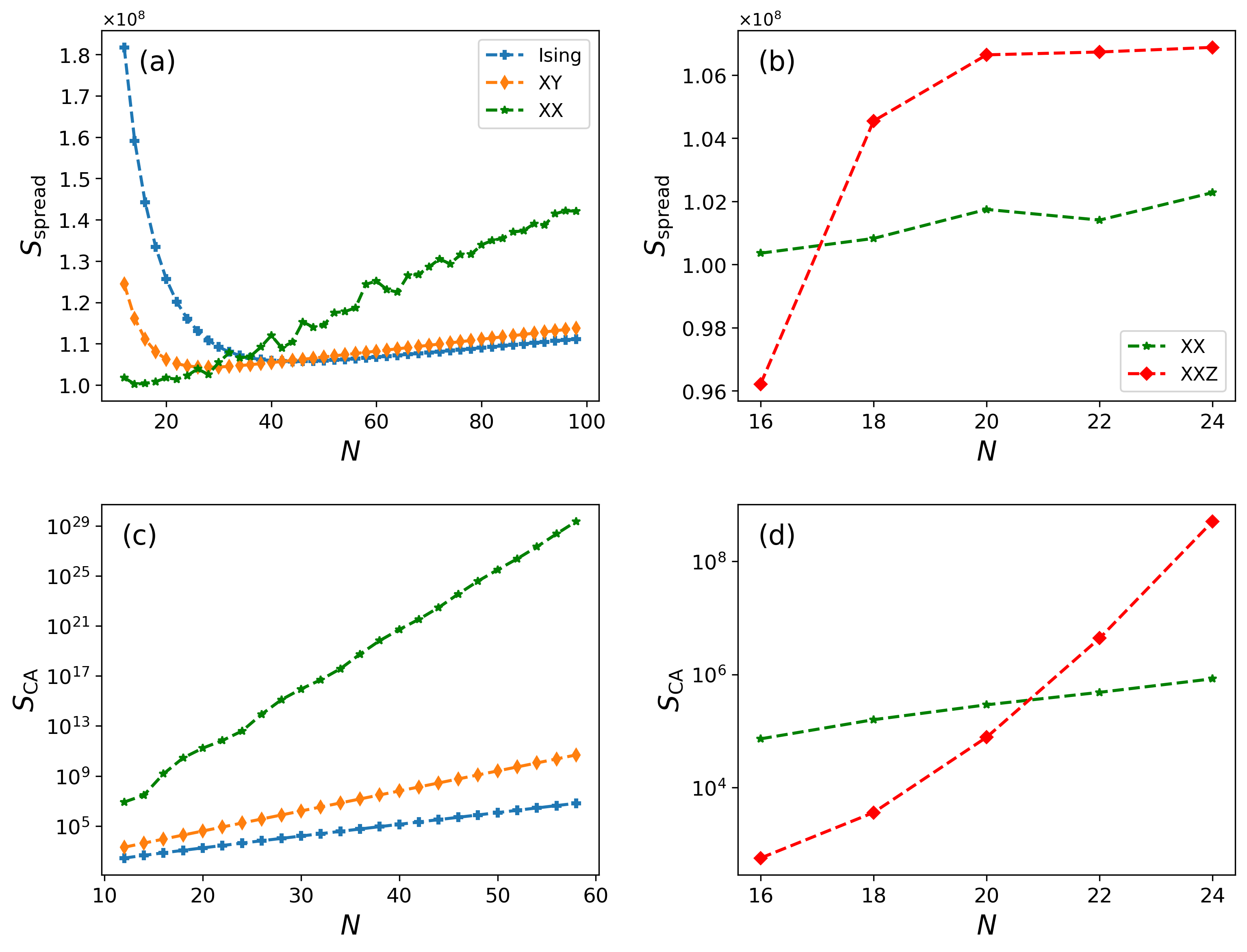}
    \caption{Finite-size bounds on the number of measurement shots needed to estimate the global fidelity kernel of Eq.~\eqref{eq:global} via the SWAP test. The fidelity-per-site kernel of Eq.~\eqref{eq:fid_per_site} is not used in these shot-count estimates.
    The number of measurement shots is evaluated using the spread bound in Eq.~\eqref{eq:Sspread-main}, obtaining the estimate $S_{\mathrm{spread}}$, and the concentration-avoidance bound  in Eq.~\eqref{eq:SCA-main}, yielding the estimate $S_{\mathrm{CA}}$. 
    The training windows are the same as in Fig.~\ref{fig:svm_combined}: for Ising/XY/XX we sample $h\in[0.7,0.95]\cup[1.05,1.3]$ at $\Delta=0$ (Ising $\gamma=1$, XY $\gamma=0.5$, XX  $\gamma=10^{-3}$), while for XXZ we sample $\Delta\in[0.35,0.45]\cup[0.55,0.65]$ at $h=0$ and $\gamma=10^{-3}$.
    (a) $S_{\mathrm{spread}}$ versus $N$ for the free-fermion XY family (Ising/XY/XX).
    (b) $S_{\mathrm{spread}}$ comparison between XX and XXZ for $N=16$--$24$.
    (c) $S_{\mathrm{CA}}$ versus $N$ for the XY family (log scale).
    (d) $S_{\mathrm{CA}}$ comparison between XX and XXZ for $N=16$--$24$ (log scale).
    These finite-size estimates are obtained by putting 
    $\varepsilon=10^{-3}$, $P_{\mathrm{spread}}=0.99$, 
     $\varepsilon_{\mathrm{CA}}=10^{-3}$, and $P_{\mathrm{CA}}=0.99$.}
    \label{fig:shot_bounds}
\end{figure*}

A second, independent difficulty arises when the off-diagonal
fidelities themselves become globally small, as occurs for many-body
ground states across a phase boundary at large $N$. In that regime,
resolving the spread of entries is no longer sufficient: one must
also be able to determine the representative scale $k_{\mathrm{repr}}$
with controlled relative accuracy. To see why, we recall from
Eq.~\eqref{eq:swap_var} that for small $k_{i,j}$ the standard
deviation of the SWAP-test estimator approaches $1/\sqrt{S}$,
independently of $k_{i,j}$. A signal of order $k_{\mathrm{repr}}$
is then buried in noise unless $S$ is large enough to satisfy
$1/\sqrt{S} \ll k_{\mathrm{repr}}$. If this condition fails, all
off-diagonal entries appear equally small and the Gram matrix loses
its discriminative power entirely. We formalize this as a
relative-resolution requirement with confidence level
$P_{\mathrm{CA}}$ and fractional tolerance $\varepsilon_{\mathrm{CA}}$,
demanding that the deviation of the estimator from its mean
$\mathbb{E}[\hat{k}_{i,j}]=k_{i,j}$ remain below
$\varepsilon_{\mathrm{CA}}\,k_{\mathrm{repr}}$ with confidence
$P_{\mathrm{CA}}$,
\begin{equation}
\Pr\!\left(|\hat{k}_{i,j} - k_{i,j}| \ge
\varepsilon_{\mathrm{CA}}\,k_{\mathrm{repr}}\right)
\le 1 - P_{\mathrm{CA}} \, .
\label{eq:CA_req}
\end{equation}
Applying Chebyshev's inequality, Eq.~\eqref{eq:chebyshev}, to
$X=\hat{k}_{i,j}$ with threshold
$t=\varepsilon_{\mathrm{CA}}\,k_{\mathrm{repr}}$, and again using the
variance from Eq.~\eqref{eq:swap_var}, gives
\begin{equation}
\Pr\!\left(|\hat{k}_{i,j} - k_{i,j}| \ge
\varepsilon_{\mathrm{CA}}\,k_{\mathrm{repr}}\right)
\le
\frac{1-k_{i,j}^2}{S\,\varepsilon_{\mathrm{CA}}^2\,k_{\mathrm{repr}}^2} \, .
\label{eq:chebyshev_CA}
\end{equation}
Comparing Eq.~\eqref{eq:chebyshev_CA} with the requirement in
Eq.~\eqref{eq:CA_req}, it is sufficient to impose
\begin{equation}
\frac{1-k_{i,j}^2}{S\,\varepsilon_{\mathrm{CA}}^2\,k_{\mathrm{repr}}^2}
\le 1-P_{\mathrm{CA}} \, ,
\end{equation}
which, solved for $S$ and with $k_{i,j}$ replaced by the
representative value $k_{\mathrm{repr}}$, as in the derivation of
Eq.~\eqref{eq:Sspread-main}, gives the concentration-avoidance (CA)
bound
\begin{equation}
S_{\mathrm{CA}} \ge
\frac{1 - k_{\mathrm{repr}}^2}
{(1 - P_{\mathrm{CA}})\,\varepsilon_{\mathrm{CA}}^2\,
k_{\mathrm{repr}}^2} \, .
\label{eq:SCA-main}
\end{equation}
Since $1-k_{i,j}^{2}\le 1$ for every entry, replacing the numerator
by unity yields a fully rigorous worst-case version of the same
bound, which coincides with Eq.~\eqref{eq:SCA-main} in the
concentrated regime $k_{\mathrm{repr}}\ll 1$ that concerns us here.
For $k_{\mathrm{repr}}\ll 1$, Eq.~\eqref{eq:SCA-main} reduces to
$S_{\mathrm{CA}} \propto k_{\mathrm{repr}}^{-2}$ and is recognized
as the quantitative form of the heuristic requirement
$1/\sqrt{S} \ll k_{\mathrm{repr}}$ introduced above: the shot-noise
floor must be pushed below the representative fidelity scale itself.
The two bounds therefore probe complementary failure modes of the
estimated Gram matrix. The spread bound of
Eq.~\eqref{eq:Sspread-main} diverges when the ensemble spread
collapses, $\Delta_{\mathrm{ensemble}}\to 0$, irrespective of the
overall fidelity scale, whereas the CA bound diverges when the
overall scale collapses, $k_{\mathrm{repr}}\to 0$, irrespective of
the spread. As both requirements must hold simultaneously, the
experimentally relevant shot count is set by the larger of the two.
The physical origin of exponentially small values of
$k_{\mathrm{repr}}$, and the resulting magnitude of
$S_{\mathrm{CA}}$ reported in Fig.~\ref{fig:shot_bounds}, are
discussed in Sec.~\ref{results}.

Lastly, we want to emphasize that our concentration-avoidance requirement differs from the concentration criterion of Ref.~\cite{miroszewski2024search}. There, avoidance of concentration is imposed through a logarithmic condition tailored to the exponential kernel concentration of the ZZ feature map. Here, by contrast, Eq.~\eqref{eq:CA_req} is a relative-resolution requirement obtained directly from Chebyshev's inequality applied to the SWAP-test estimator: it demands that the estimator deviation stay below a fixed fraction of the representative scale $k_{\mathrm{repr}}$, which yields the algebraic form $S_{\mathrm{CA}}\propto k_{\mathrm{repr}}^{-2}$ rather than a logarithmic one. The two criteria therefore address the same qualitative difficulty, the collapse of the overall kernel scale, but through distinct estimators and with distinct scaling the algebraic form is the one appropriate to the fidelity kernel measured by the SWAP test, whose variance is fixed by Eq.~\eqref{eq:swap_var}.

\section{Results}
\label{results}
In this section,  we discuss the results of our resource-estimation analysis and report the main numerical findings.
We compare the bounds on the number of measurement shots across the four models under study, and highlight how the required measurement resources scale with the system size and with the choice of control parameter.

\subsection{Free-fermion XY family}

In Fig.~\ref{fig:shot_bounds}(a), we report the required number of shots $S_{\mathrm{spread}}$ to estimate the kernel entries $\hat k_{i,j}$ within the spread criterion.
 For all panels in Fig.~\ref{fig:shot_bounds}, the kernel entries \(k_{ij}\) are those of the global fidelity kernel defined in Eq.~\eqref{eq:global}. 
  This is the quantity directly estimated by the SWAP test through Eq.~\eqref{eq:K_swap}. We recall that the variance entering the shot bounds is 
$  \mathrm{Var}(\hat{k}_{ij})=\frac{1-k_{ij}^{2}}{S}$  which applies to the directly measured global-fidelity estimator. Using the fidelity-per-site kernel in a resource analysis would require a separate treatment, because it corresponds to a nonlinear transformation of the measured global fidelity.

We observe that $S_{\mathrm{spread}}$ can be very large for small system sizes, which we mainly attribute to finite-size effects: for small $N$ the kernel ensemble does not yet display a stable separation between the two parameter regions used in the training set.
As $N$ increases, ground states taken from opposite sides of the transition become progressively more distinguishable, since their overlaps decrease, and the kernel distribution becomes more spread out. As a result, the bound predicts a reduction in the required number of shots up to an intermediate size.
Interestingly, for larger $N$ the required shots increase again. This non-monotonic behavior is due to a concentration of many off-diagonal kernel values near zero, which reduces the ensemble spread $\Delta_{\mathrm{ensemble}}$. Since the bound requires $|\hat k_{i,j}-\mathbb{E}[\hat k_{i,j}]| < \varepsilon\,\Delta_{\mathrm{ensemble}}$, a smaller $\Delta_{\mathrm{ensemble}}$ implies a larger number of shots, scaling as $S_{\mathrm{spread}} \propto 1/\Delta_{\mathrm{ensemble}}^{2}$.
We also find that reducing the anisotropy parameter $\gamma$ (i.e.~by passing from Ising to XY, and then to the XX limit) significantly increases the resource cost. This hierarchy ($\mathrm{Ising} < \mathrm{XY} < \mathrm{XX}$) is driven by the sensitivity of the ground state wavefunction to parameter changes.
To understand the origin of this behavior, we compute the rate at which the ground state rotates in Hilbert space. The ground state is determined by the Bogoliubov angles $\theta_q(h)$, which satisfy $\tan\!\big(2\theta_q\big) = w_q$ with $w_q = \frac{\gamma \sin q}{h - \cos q}$ (see Appendix~\ref{app:bogoliubov_fidelity} for more details).
Applying the chain rule $\partial_h \theta_q = \frac{1}{2} \frac{1}{1+w_q^2} \frac{\partial w_q}{\partial h}$, we find:
\begin{align}\label{eq:theta_derivative}
\frac{\partial \theta_q}{\partial h} &= \frac{1}{2} \frac{(h-\cos q)^2}{(h-\cos q)^2 + \gamma^2 \sin^2 q} \left[ \frac{-\gamma \sin q}{(h-\cos q)^2} \right] \nonumber \\
&= -\frac{1}{2} \frac{\gamma \sin q}{E_q^2} \, , 
\end{align}
where $E_q^2 = (h-\cos q)^2 + \gamma^2 \sin^2 q$ is the squared energy gap and $q$ is momentum mode.
Eq.~\eqref{eq:theta_derivative} reveals that the wavefunction sensitivity is inversely proportional to the square of the energy gap, scaled by the anisotropy $\gamma$. 

Before analyzing each model, we make explicit how this sensitivity controls the exponential decay of the fidelity, and hence the shot bounds.
We recall that for the free fermion family the fidelity factorizes over momentum modes, see Eq.~\eqref{eq:app_fidelity_product},
\begin{equation}
F(h_a,h_b)=\prod_{q>0}\cos^{2}\!\big(\Delta\theta_q\big) \, ,
\qquad \Delta\theta_q \equiv \theta_q(h_b)-\theta_q(h_a) \, .
\label{eq:F_product_recall}
\end{equation}
Because each factor obeys $0\leq \cos^{2}(\Delta\theta_q)\leq1$, a product of $N/2$
such factors is exponentially small in $N$. Writing
$\theta_q(h)=\tfrac12\arg\!\left[\epsilon_q(h)+i\Lambda_q\right]$, with
$\epsilon_q(h)=h-\cos q$ and $\Lambda_q=\gamma\sin q$, each factor follows in
closed form,
\begin{equation}\label{eq:mode_overlap_exact}
\cos^{2}\!\big(\Delta\theta_q\big)=\frac{1}{2}
\left[1+\frac{\epsilon_q(h_a)\,\epsilon_q(h_b)+\Lambda_q^{2}}
{E_q(h_a)\,E_q(h_b)}\right] ,
\end{equation}
with $E_q(h)=\sqrt{\epsilon_q(h)^{2}+\Lambda_q^{2}}$, so that taking the log gives
\begin{equation}\label{eq:fid_sus}
    \ln F = \sum_{q>0}\ln\cos^{2}\!\big(\Delta\theta_q\big) \equiv -aN .
\end{equation}
Since the sum runs over $N/2$ modes, the decay rate $a$ is intensive and the global fidelity obeys $F\sim e^{-aN}$ for all three models under consideration. What distinguishes the models is therefore not whether the fidelity decays exponentially but the rate $a$. Equation~\eqref{eq:mode_overlap_exact} is exact for any field separation. As a function of $h$, the angle $\theta_q$ interpolates between two constant values across a crossover of width $\Lambda_q$ centred on the gap-closing point $h=\cos q$. A mode therefore has small $\Delta\theta_q$ unless that point falls inside the sampled window while $\Lambda_q$ is small compared with $|\epsilon_q|$
at the endpoints. Away from that exception,
$-\ln\cos^{2}(\Delta\theta_q)\simeq(\Delta\theta_q)^{2}$, which with
Eq.~\eqref{eq:theta_derivative} and $\delta h=h_b-h_a$ gives
\begin{equation}\label{eq:fid_sus_pert}
    a \simeq \frac{(\delta h)^{2}}{N}
    \sum_{q>0}\left(\frac{\gamma\sin q}{2E_q^{2}}\right)^{2} .
\end{equation}
This is the regime of cases (i) and (ii) below. It does not extend to the XX limit,
where $\epsilon_q$ changes sign inside the window while $|\epsilon_q|$ exceeds
$\Lambda_q$ at its endpoints, so that $\Delta\theta_q$ approaches the zero of the
cosine and no expansion about $\Delta\theta_q=0$ applies. There we use Eq.~\eqref{eq:mode_overlap_exact} directly. Combined with $k_{\mathrm{repr}}\sim e^{-aN}$ and Eq.~\eqref{eq:SCA-main} this gives directly 
\begin{equation}
    S_\mathrm{CA}\sim e^{2aN},
\label{eq:SCA_exp}
\end{equation}
so any increase in $a$ translates into exponentially larger shot cost. We now will examine the rate $a$ for each model 

\textit{(i) Ising Limit ($\gamma=1$):} The model has a discrete $\mathbb{Z}_2$ symmetry. Away from $h_c$ the gap is finite, so $|\partial_h\theta_q|$ stays bounded.  For the modes whose gap-closing point falls inside the sampled window, $\Lambda_q$ still exceeds $|\epsilon_q|$ at the endpoints, so the crossover is broad, no mode is fully rotated, the small-angle limit of Eq.~\eqref{eq:mode_overlap_exact} applies, every term of Eq.~\eqref{eq:fid_sus_pert} is finite, and $a$ takes its smallest value among the three models. Through $S_{\mathrm{CA}}\sim e^{2aN}$, this small rate translates into the lowest shot cost,  and the kernel ensemble retains a larger spread.

\textit{(ii) XY Model ($\gamma=0.5$):} The symmetry is still $\mathbb{Z}_2$, preserving a finite gap.  However, each term of Eq.~\eqref{eq:fid_sus_pert} carries a factor $\gamma^{2}\sin^{2}q$ in its numerator and $E_q^{4}$ in its denominator, and while $\Lambda_q$ still exceeds $|\epsilon_q|$ for most of the modes that cross the gap-closing point, the quantity $E_q^{2}$ itself scales with $\gamma^{2}$, so the term behaves as $\gamma^{2}/(\gamma^{2})^{2}=1/\gamma^{2}$. Reducing $\gamma$ from $1$ to $0.5$ therefore increases the decay rate by roughly a factor of four relative to Ising. This $1/\gamma^{2}$ scaling holds only while the crossover of $\theta_q$ remains broad, and it does not extend to the XX limit.  This sharper response leads to faster orthogonality decay compared to the Ising case,  a larger rate $a$, and hence a consistently higher shot cost through $S_{\mathrm{CA}}\sim e^{2aN}$. 

\textit{(iii) XX Limit ($\gamma \to 0$):} This regime represents a fundamental change in the universality class. As $\gamma \to 0$, the Hamiltonian restores a continuous $U(1)$ rotation symmetry. This forces the energy gap to close ($E_q \to 0$ at the critical momentum defined by $h - \cos q_c = 0$).
Inspecting Eq.~\eqref{eq:theta_derivative} at resonance ($h \approx \cos q$), the derivative diverges as:
\begin{equation}
\left| \frac{\partial \theta_q}{\partial h} \right| \approx \frac{1}{2 \gamma \sin q} \to \infty \, .
\end{equation}

On the finite-$N$ momentum grid, however, no sampled field lies within $\Lambda_q$
of a gap-closing point, so it is not this divergence that the kernel resolves.
What changes is that the crossover of $\theta_q$, broad in cases (i) and (ii), now becomes sharp: as $\gamma\to0$ its width $\Lambda_q$ vanishes while $|\epsilon_q|$ stays finite at the sampled fields. For every mode whose gap-closing point falls inside the sampled window, $\epsilon_q$ changes sign while $|\epsilon_q|$ exceeds $\Lambda_q$ at both
endpoints, so $\Delta\theta_q$ approaches the zero of the cosine and
Eq.~\eqref{eq:mode_overlap_exact} reduces to
\begin{equation}\label{eq:switch_factor}
\cos^{2}\!\big(\Delta\theta_q\big)\simeq
\left[\frac{\Lambda_q}{2}\left(\frac{1}{|\epsilon_q(h_a)|}
+\frac{1}{|\epsilon_q(h_b)|}\right)\right]^{2} ,
\end{equation}
so that each such mode contributes $2\ln(1/\gamma)$ to $-\ln F$. The number of
such modes grows in proportion to $N$, so the rate $a$ remains intensive but now
increases logarithmically as the $U(1)$ limit is approached, rather than
algebraically as in case (ii). This is the qualitative difference: in the gapless
$U(1)$ phase the suppression is set by how completely each mode is rotated, not by
how fast it rotates, and a finite parameter step carrying a mode through its
gap-closing point rotates it by the maximum amount available.

\begin{figure}[t]
    \centering
    \includegraphics[width=\linewidth]{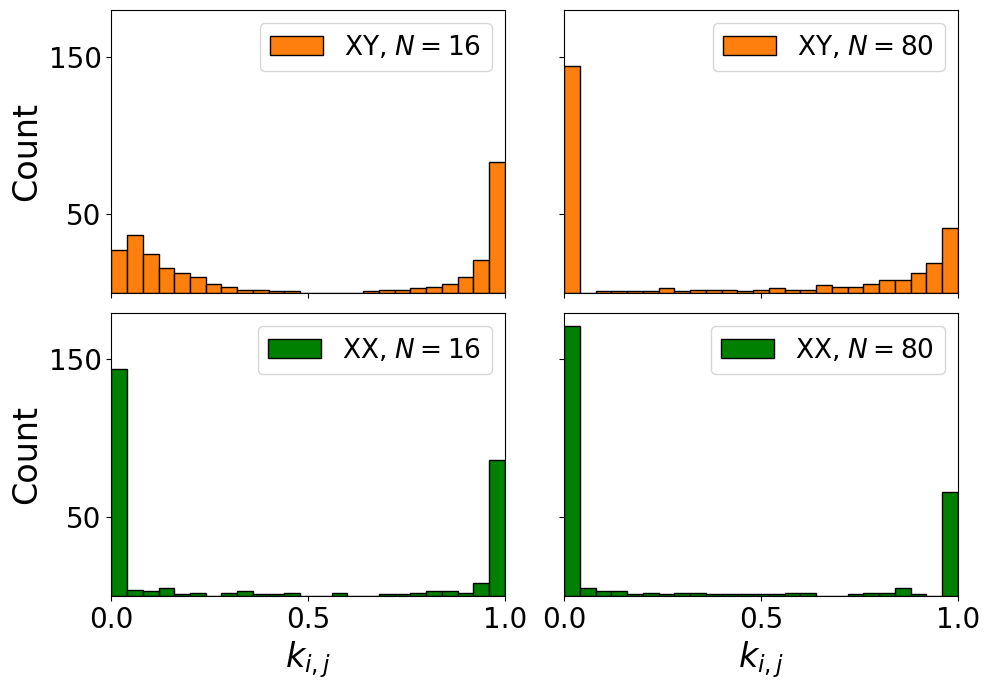}
    \caption{Histogram of the fidelity-kernel entries $k_{i,j}$ (upper-triangular part) for the XY and XX models at two system sizes. The left column shows $N=16$ (top: XY, bottom: XX), while the right column shows $N=80$ (top: XY, bottom:XX).}
    \label{fig:count_zero}
\end{figure}

This leads to the kernel values concentrating near zero so rapidly that the cost rises monotonically for all observed $N \ge 16$, lacking the non-monotonic minimum seen in case of Ising and XY.
This mechanism is further illustrated in Fig.~\ref{fig:count_zero}.
Since kernel entries concentrate near zero in this regime, it is useful to focus on the CA-type shot requirement in Eq.~\eqref{eq:SCA-main}, in order to avoid such concentration.
Accordingly, Fig.~\ref{fig:shot_bounds}(c) reports the number of shots needed to avoid concentration. 
When the kernel values concentrate near zero ($k_{\mathrm{repr}}\to 0$), the required number of shots blows up since $S_{\mathrm{CA}} \propto 1/k_{\mathrm{repr}}^{2}$.

It is useful to read this relation backwards. With the parameters used in Fig.~\ref{fig:shot_bounds}, namely $\varepsilon_{\mathrm{CA}}=10^{-3}$ and $P_{\mathrm{CA}}=0.99$, Eq.~\eqref{eq:SCA-main} in the concentrated regime $k_{\mathrm{repr}}\ll1$ becomes $S_{\mathrm{CA}}\simeq10^{8}/k_{\mathrm{repr}}^{2}$, or equivalently $k_{\mathrm{repr}}\simeq10^{4}/\sqrt{S_{\mathrm{CA}}}$. A shot count of $10^{29}$ therefore corresponds to resolving a representative fidelity of order $10^{-11}$. The extreme values reported in Fig.~\ref{fig:shot_bounds}(c,d) are in this sense a direct statement about the magnitude of the cross-phase overlaps rather than an artifact of the bound.

Importantly, this rapid growth of the shot count should not be interpreted as a failure of the learning protocol itself. 
Rather, it quantifies the experimental cost of faithfully estimating the fidelity kernel through direct SWAP-test measurements. 
In this sense, the divergence of $S_{\mathrm{CA}}$ provides a practical criterion for identifying the regimes in which fidelity-kernel learning becomes measurement limited, 
indicating where alternative kernel constructions or more efficient fidelity-estimation techniques may be required.

\subsection{XXZ chain}

We treated the XXZ case separately because, unlike the XY family, we do not have access to a simple closed-form analytical expression for the ground-state wave function in the interacting regime. Therefore, we rely on exact numerical diagonalization,
\begin{equation}
\ket{\psi(\Delta)}=\arg\min_{\|v\|=1}\bra{v}H(\Delta)\ket{v}.
\label{Exact_Diag}
\end{equation}

For the largest size considered below ($N=50$), where exact diagonalization is not feasible, the ground states and their overlaps are obtained with DMRG matrix-product-state simulations~\cite{RevModPhys.77.259}.

In the XXZ limit ($\gamma=0$) at zero field ($h=0$), the model exhibits critical points at $\Delta=\pm 1/2$.
$\Delta=-1/2$ corresponds to a Berezinskii--Kosterlitz--Thouless (BKT) transition, while $\Delta=+1/2$ marks the ferromagnetic boundary~\footnote{These values for the critical points follows from our sign convention in Eq.~\eqref{eq:hamiltonian}. Note that this differs from the more common notation found in literature \cite{Giamarchi2004Q1D,CabraPujol2004,MikeskaKolezhuk2004,Wang2010,Pradhan2025Oddities}, where the critical values are
$\tilde{\Delta}=\pm 1$ . The relation with our notation is
$\Delta=-\tilde{\Delta}/2$.}.
Here we focus on the ferromagnetic transition, and accordingly we train the SVM in parameter windows around $\Delta= 1/2$.
 
The BKT transition at
$\Delta=-1/2$ is treated separately in
Sec.~\ref{sec:xxz_bkt}.

\begin{figure*}[t]
    \centering
    \includegraphics[width=\linewidth]{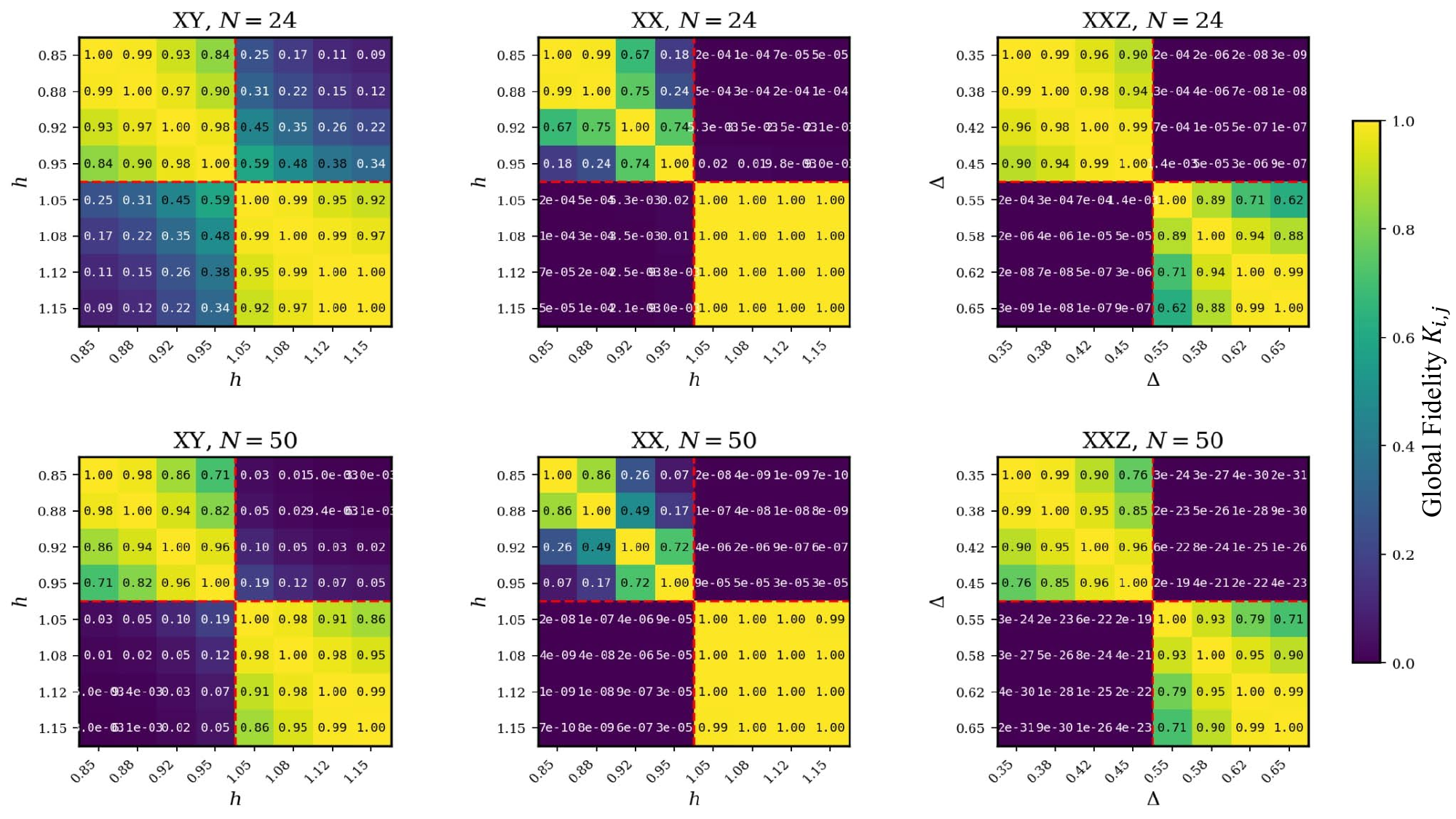}
    \caption{Global fidelity kernel $k_{i,j}$ for the XY, XX, and XXZ models at $N=24$ (top row) and $N=50$ (bottom row), with each entry printed explicitly. Model parameters are those of Fig.~\ref{fig:shot_bounds} (XY: $\gamma=0.5$, XX: $\gamma=10^{-3}$ XXZ: $\gamma=10^{-3}$, $h=0$), and the axes label representative training points taken from the two windows on opposite sides of the transition ($h$ for XY/XX, $\Delta$ for XXZ). The red dashed lines separate the two windows, so that the off-diagonal blocks contain the cross-phase entries. XY and XX entries are computed from the closed-form fidelity of Eq.~\eqref{eq:app_fidelity_product} XXZ entries are computed by exact diagonalization at $N=24$ and by DMRG at $N=50$. The cross-phase(off diagonal) blocks collapse at model-dependent rates: at $N=50$ they span $\sim\!10^{-1}$--$10^{-3}$ for XY, $\sim\!10^{-4}$--$10^{-10}$ for XX, and $\sim\!10^{-19}$--$10^{-31}$ for XXZ, the smallest values approaching the limits of double-precision arithmetic.}
    \label{fig:heatmap_zoom}
\end{figure*}

Figure~\ref{fig:shot_bounds}(b) compares the XXZ chain with the XX model for system sizes $N=16,\dots,24$, around the $\Delta=1/2$ transition.
Within the XY family, the XX regime already shows the largest measurement shot cost. Here we observe that the interacting XXZ model requires even more shots according to the spread bound.
This trend is consistent with Fig.~\ref{fig:shot_bounds}(d): once $N$ is large enough, such that finite-size effects are reduced, the number of shots for the XXZ chain is larger than for the XX model within the same $N$ range.
Due to limited computational resources, we restrict to $N \le 24$ in Fig.~\ref{fig:shot_bounds}, but we expect this separation to persist at larger system sizes. 
This expectation is confirmed at the level of individual kernel entries at $N=50$ in Fig.~\ref{fig:heatmap_zoom}, as discussed below.

A natural interpretation is that, unlike the XY chains, the XXZ model is genuinely interacting.
Interaction modifies the distribution of kernel entries and can reduce the ensemble spread, thereby increasing $S_{\mathrm{spread}}$.
Concerning symmetries, the XX chain has a $U(1)$ symmetry.
By contrast, the XXZ chain 
becomes fully $SU(2)$-symmetric at the isotropic point 
$\Delta= 1/ 2$ for $h=0$.
In terms of our resource bounds, the $SU(2)$ symmetry tends to reshape the distribution of kernel entries by increasing concentration and reducing the effective spread $\Delta_{\mathrm{ensemble}}$ (and driving $k_{\mathrm{repr}}$ toward smaller values), which directly amplifies both $S_{\mathrm{spread}}$ and $S_{\mathrm{CA}}$ through Eqs.~\eqref{eq:Sspread-main} and \eqref{eq:SCA-main}. This effect is absent in the free-fermion XX chain, where the $U(1)$ symmetry is present but the ground-state manifold is simpler and admits a non interacting description.

This interpretation can be made quantitative. For the free-fermion family, the decay rate $a$ of Eq.~\eqref{eq:fid_sus} was obtained from the mode factorization of the fidelity, Eq.~\eqref{eq:F_product_recall}. For the XXZ chain this factorization no longer holds, because the $\Delta\,\sigma_i^z\sigma_{i+1}^z$ interaction couples different momentum modes. The fidelity, however, still decays exponentially, $F\sim e^{-aN}$, and the rate can be obtained from standard perturbation theory. Writing $H(\Delta+\delta\Delta)=H(\Delta)+\delta\Delta\,\hat{V}$ with $\hat{V}=\partial_\Delta H=-\sum_i\sigma_i^z\sigma_{i+1}^z$, and denoting by $\{\ket{n},E_n\}$ the eigenstates of $H(\Delta)$, the overlap between neighboring ground states reads
\begin{equation}
-\ln F \approx (\delta\Delta)^{2}\sum_{n\neq0}\frac{|\bra{n}\hat{V}\ket{0}|^{2}}{(E_n-E_0)^{2}}
\;\equiv\; \chi_F\,(\delta\Delta)^{2}\,N \, ,
\label{eq:kubo_sum}
\end{equation}
where $\chi_F$ denotes the fidelity susceptibility per site~\cite{Gu2010Review}, so that the decay rate is $a=\chi_F\,(\delta\Delta)^{2}$, in complete analogy with Eq.~\eqref{eq:fid_sus_pert}. Indeed, Eq.~\eqref{eq:kubo_sum} is the general form of which Eq.~\eqref{eq:fid_sus_pert} is the closed-form free-fermion evaluation: there, the sum over excited states reduces to a sum over momentum modes, and the matrix elements are fixed by the Bogoliubov-angle sensitivity of Eq.~\eqref{eq:theta_derivative}.

Equation~\eqref{eq:kubo_sum} also makes transparent why the XXZ rate exceeds the XX one: the sum is large when the energy denominators are small, and when the perturbation couples the ground state strongly to excited states. Both effects are enhanced in the XXZ chain. First, near $\Delta=1/2$ the model approaches the fully $SU(2)$-symmetric Heisenberg point, where the enlarged symmetry produces additional low-lying states and the relevant excitation gap closes, in analogy with the $U(1)$ gap closing of the XX limit this reduces the denominators. Second, while in the free-fermion family the perturbation $\partial_h H=-\sum_i\sigma_i^z$ is a one-body operator that connects the ground state only to a restricted set of two-quasiparticle excitations, the bond operator $\sum_i\sigma_i^z\sigma_{i+1}^z$ acts on a genuinely interacting ground state and couples it to a much larger set of many-body excitations, enhancing the numerators. Acting in the same direction, the two effects give $a_{\mathrm{XXZ}}>a_{\mathrm{XX}}$ at equal $N$ and, through Eq.~\eqref{eq:SCA_exp}, an exponentially larger shot cost, consistent with the behavior observed in Fig.~\ref{fig:shot_bounds}(b,d).

The microscopic origin of this hierarchy is displayed in Fig.~\ref{fig:heatmap_zoom}, which shows the Gram matrices of the three models at $N=24$ and $N=50$ with each entry printed explicitly. Within the diagonal blocks, which collect same-phase pairs, the entries closest to the diagonal remain of order unity for all models and both sizes: states separated by a small parameter step inside a single training window stay strongly overlapping, so the phase structure of the kernel is well preserved and the classification task itself remains well posed. The same-phase entries do decay across the full width of a window, but far more slowly than the cross-phase ones. The collapse is confined to the cross-phase blocks, and it proceeds at strongly model-dependent rates, growing sharply from XY to XX to XXZ in agreement with the decay rates of Eqs.~\eqref{eq:fid_sus} and \eqref{eq:kubo_sum}. Since $k_{\mathrm{repr}}$ is dominated by these cross-phase entries as $N$ grows, their collapse is what drives the concentration-avoidance cost through $S_{\mathrm{CA}}\propto k_{\mathrm{repr}}^{-2}$. This separation shows that the growing resource cost is not a failure of classification, since the SVM continues to separate the phases from the well-resolved block structure. Rather, it reveals a measurement limitation: resolving exponentially small cross-phase fidelities at fixed relative accuracy is what drives $S_{\mathrm{CA}}$ to the extreme values reported here. This provides a concrete physical reading of the symmetry-resource hierarchy $\mathbb{Z}_2 < U(1) < SU(2)$: increasing symmetry closes the relevant gap and accelerates the orthogonalization of cross-phase ground states, and the measurement cost grows accordingly.

\subsection{BKT transition in the XXZ chain} 
\label{sec:xxz_bkt}

\begin{figure}
    \centering
    \includegraphics[width=\linewidth]{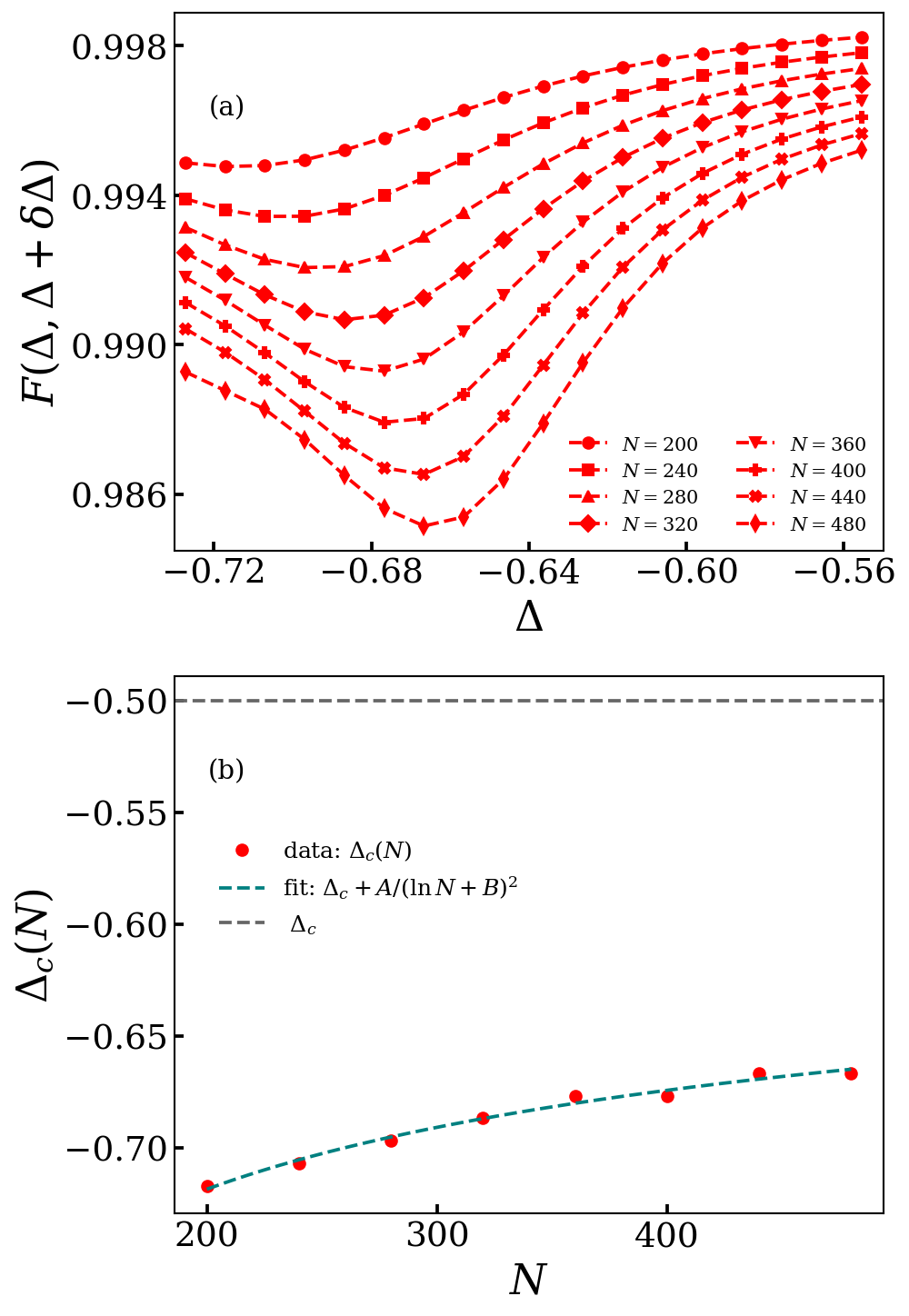}
    \caption{(a) Ground-state fidelity $F(\Delta,\Delta+\delta\Delta)$ between neighbouring values of the control parameter \(\Delta\), computed with $\delta\Delta=0.005$, for system sizes $N=200$--$480$. The fidelity feature appears away from the thermodynamic BKT point at finite $N$, showing the strong finite-size drift of the BKT transition. (b) Finite-size drift of the pseudo-critical points $\Delta_c(N)$ extracted from the fidelity data, together with the BKT-type extrapolation toward the expected critical point $\Delta_c=-0.5$. }
    \label{fig:xxz_bkt}
\end{figure}

The BKT transition is one of the most difficult phase transitions to detect, both analytically and numerically. The main reason is that the ground state changes very smoothly near the transition, and there is no simple local order parameter which becomes nonzero in the usual way. Because of this, fidelity-based quantities may show only a broad minimum, a weak feature, or a peak of the fidelity susceptibility which is shifted away from the true critical point. This issue has also been discussed in the literature, where the apparent transition point obtained from fidelity or fidelity susceptibility can appear inside the gapped phase and then slowly move toward the true BKT point when the system size is increased \cite{cincio2019universal,zhang2021fidelity,sun2015fidelity}.
This point is important for our supervised-learning analysis. For a BKT transition, the boundary obtained from the SVM at finite size should not be understood as the exact thermodynamic critical point. Instead, it should be understood as a finite-size pseudo-critical point. In our convention for the XXZ chain, the expected BKT point is $\Delta_c=-0.5$. We find that the nearest-neighbour fidelity develops a minimum on the gapped side of the transition, and this minimum moves slowly toward $\Delta_c=-0.5$ as the system size is increased. This is completely consistent with the known behavior of fidelity-based probes for BKT transitions. We show this behavior in Fig.~\ref{fig:xxz_bkt}(a).
The reason for this slow drift is the BKT form of the correlation-length divergence~\cite{kosterlitz2016kosterlitz},
\begin{equation}
\xi(\Delta) \sim \xi_0 \exp\!\left(\frac{c}{\sqrt{|\Delta-\Delta_c|}}\right), \label{eq:bkt_xi} 
\end{equation}
where $\xi$ is the correlation length, $\xi_0$ is a nonuniversal microscopic scale, and $c>0 $ is a nonuniversal constant. Since the correlation length diverges exponentially, very large system sizes are required before a finite-size estimator becomes close to the true thermodynamic point. By setting $\xi(\Delta_c(N))\sim N$, one obtains the usual BKT finite-size drift form
\begin{equation}\label{eq:bkt_drift} 
\Delta_c(N)=\Delta_c+\frac{A}{\big(\ln N+B\big)^2}, 
\end{equation}
where $A$ and $B$ are fitting parameters. Using Density Matrix Renormalization Group (DMRG)~\cite{RevModPhys.77.259}, we extract the finite-size pseudo-critical points up to $N=480$, as shown in Fig.~\ref{fig:xxz_bkt}(b). Fitting these points with Eq.~\eqref{eq:bkt_drift}, we obtain $\Delta_c^{\rm obtained}\approx -0.5201$, which is close to the expected critical point $\Delta_c=-0.5$. Therefore, the fidelity-based signal does detect the BKT transition, but it does so through a slowly drifting finite-size feature. This observation will also be important below when we discuss the SVM boundary: the SVM is not expected to give the exact BKT point at finite $N$, but it should follow the same finite-size transition region encoded in the fidelity kernel.
\begin{figure*}
        \centering
    \includegraphics[width=\linewidth]{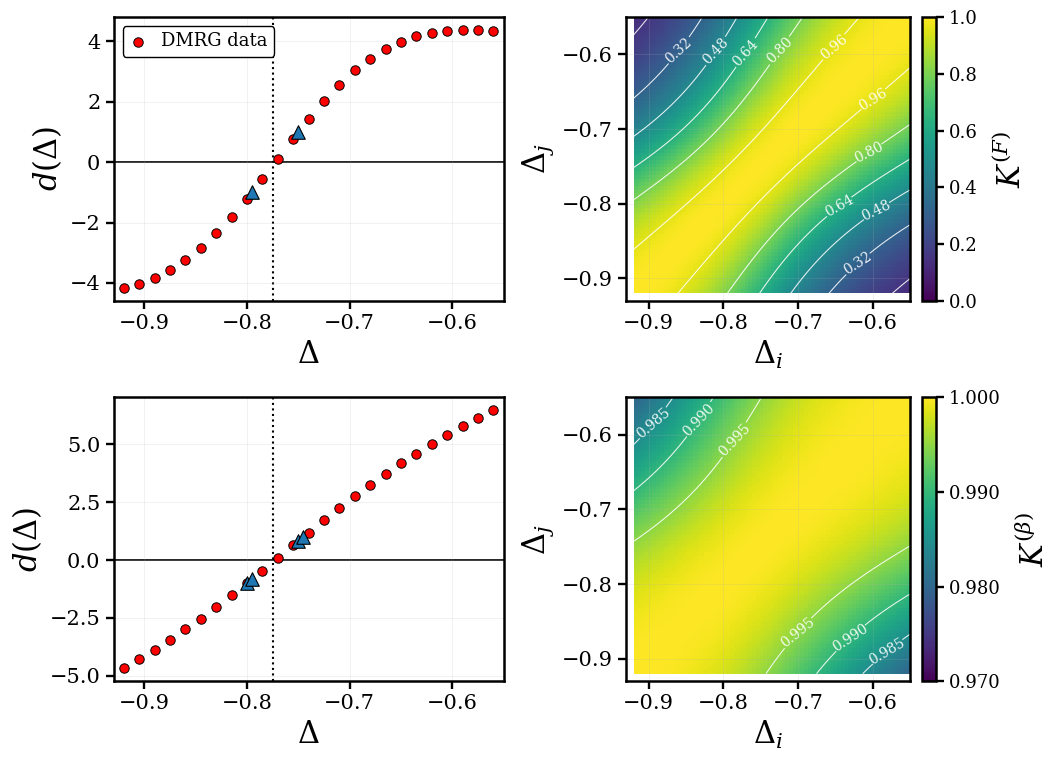}
    \caption{SVM decision-function analysis in the finite-size BKT region of the XXZ chain. Results are shown for a system size of $N=100$ at zero field, utilizing the control parameter $\Delta$ across the scanned interval $-0.90 \leq \Delta \leq -0.45$. The SVM is trained on two distinct windows, $[-0.8345, -0.7945]$ and $[-0.7545, -0.7145]$, with $16$ training points sampled from each side ($M=32$ total). The heatmap illustrates the fidelity-kernel geometry within the scanned $\Delta$ interval, while the decision-function panel displays the signed SVM output, $d(\Delta)$, derived from the same kernel. The vertical reference line marks the finite-size fidelity feature rather than the thermodynamic BKT point at $\Delta_c = -0.5$. In each case, the upper panel is computed from the global fidelity kernel of Eq.~\eqref{eq:global}, whereas the lower panel is computed from the fidelity-per-site kernel of Eq.~\eqref{eq:fid_per_site}.}
    \label{fig:BKT_d}
\end{figure*}

Having established the finite-size drift of the fidelity feature, we now apply the same SVM protocol used above for the Ising/XY/XX family to the BKT boundary of the XXZ chain. We recall that, in Sec.~\ref{SVM_train}, the training set is chosen from two parameter windows on opposite sides of the transition, and the SVM estimate is obtained from the zero crossing of the signed decision function defined in Eq.~\eqref{eq:DF}. For the BKT transition, however, this zero crossing should not be compared directly with the thermodynamic value $\Delta_c=-0.5$ at finite $N$. Since the fidelity feature itself is shifted at finite size, as shown in Fig.~\ref{fig:xxz_bkt}, the SVM boundary must be interpreted as a finite-size pseudo-critical estimator.
For system size $N=100$, we therefore train the SVM using two windows placed on the two sides of the finite-size BKT region. The decision function is then evaluated on the full scanned interval, and the SVM estimate is defined by the condition $d(\Delta_{\rm SVM})=0$. Figure~\ref{fig:BKT_d} shows the kernel geometry (heatmap) and the corresponding SVM decision function in the BKT region. The global fidelity kernel already shows a visible change of structure across the scanned interval, but the contrast is reduced by the strong suppression of many-body overlaps. The fidelity-per-site kernel removes this dominant size dependence and makes the same structure more visible. The decision function changes sign in the same region where the fidelity feature appears. 

To further characterize how the fidelity kernel guides the SVM decision boundary, we perform a shifted-window analysis. The purpose of this test is to check how stable the learned boundary is when the two training regions are moved in a controlled way around the reference point. Since the SVM decision function in Eq.~\eqref{eq:DF} is built from the kernel similarities between the test point and the training data, changing the position of the training windows provides a direct way to see whether the boundary is mainly following the chosen training configuration or whether it remains anchored by the geometry of the fidelity kernel.

\begin{figure}[htbp]
    \centering
    \includegraphics[width=0.75\linewidth]{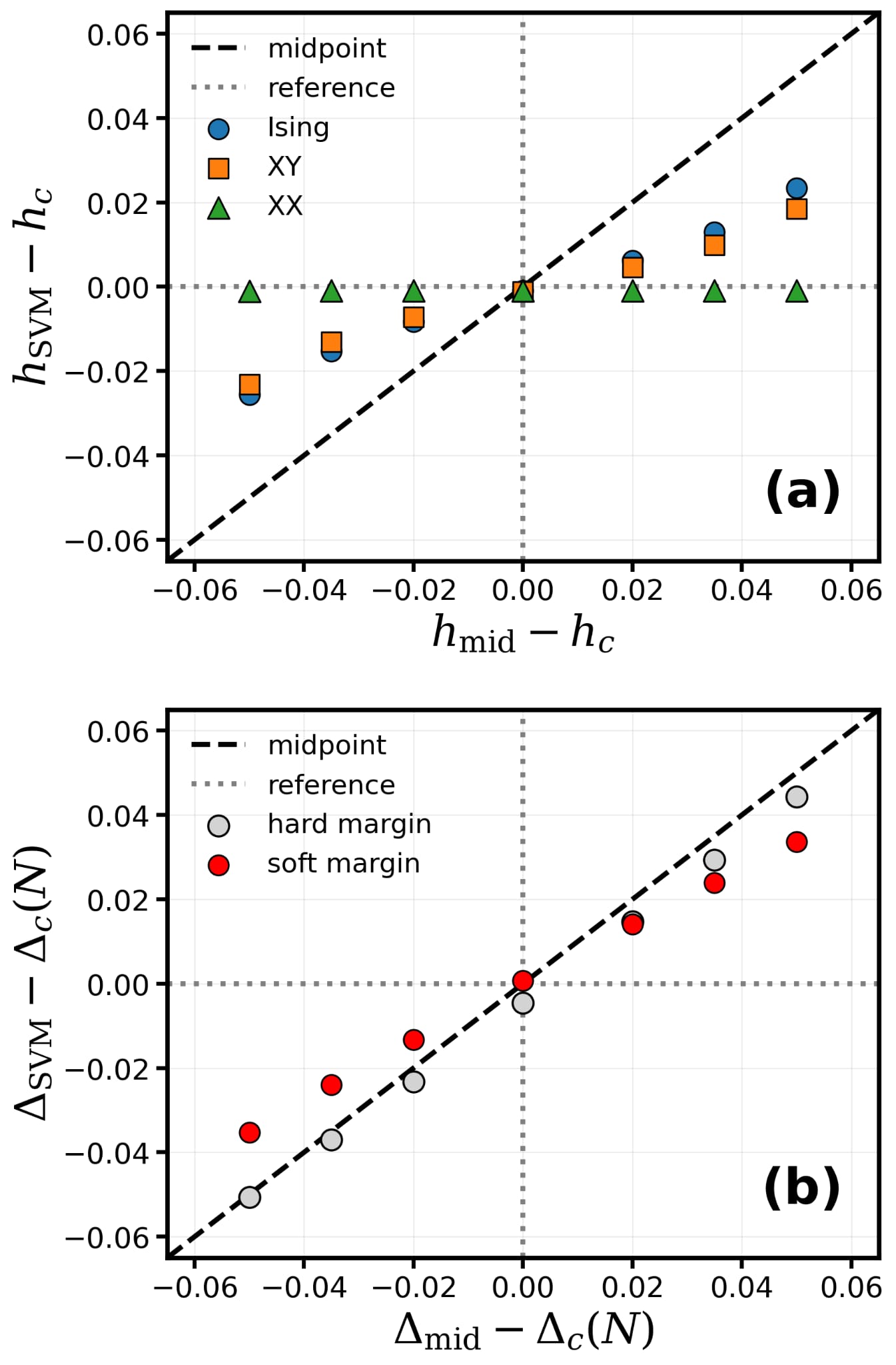}
    \caption{
    Response of the SVM decision boundary to shifted training windows. 
    The top panel (a) shows the XY-family test for a  chain length $N=50$, using the critical-field reference $h_c=1$. 
    For each shifted midpoint, $h_{\rm mid}=h_c+\delta$, the SVM is trained using two separated windows,
    $[h_{\rm mid}-0.14,h_{\rm mid}-0.06]$ and $[h_{\rm mid}+0.06,h_{\rm mid}+0.14]$,
   with $16$ training points sampled from each side ($M=32$ total).      The bottom panel (b) shows the corresponding BKT test for the $N=100$ XXZ kernel, using the finite-size reference $\Delta_c(100)=-0.7625$. 
    For each shifted midpoint, $\Delta_{\rm mid}=\Delta_c(100)+\delta$, the training windows are
    $[\Delta_{\rm mid}-0.14,\Delta_{\rm mid}-0.06]$ and
    $[\Delta_{\rm mid}+0.06,\Delta_{\rm mid}+0.14]$,
    again with $16$ training points from each side ($M=32$ total). 
    The midpoint shifts are $\delta=-0.050,-0.035,-0.020,0,0.020,0.035,0.050$. 
    In the BKT panel, the gray open circles denote the hard-margin, zero-threshold SVM boundary, while the red circles denote the cost-sensitive soft-margin estimate defined by $d(\Delta)=T$, using $T=0.25$, $C_L=0.1$, and $C_R/C_L=1.122$; these quantities are kept fixed for all window shifts. 
    The black dashed diagonal denotes complete midpoint control, while the gray dotted lines mark the corresponding reference point. }
    \label{fig:sv_midpoint_sensitivity}

\end{figure}

The result is shown in Fig.~\ref{fig:sv_midpoint_sensitivity}. The horizontal axis gives the shift of the training-window midpoint with respect to the reference point, while the vertical axis gives the corresponding shift of the SVM boundary. The black dashed diagonal represents the midpoint response. Therefore, points close to this diagonal indicate that the SVM boundary moves almost one-to-one with the training midpoint. On the other hand, points close to the horizontal gray dotted line indicate that the SVM boundary remains close to the reference critical point even when the training windows are shifted.

For the XY family, the SVM boundary stays much closer to the reference line than to the midpoint line. This shows that the fidelity kernel contains a clear geometric signature of the transition. Among the three cases, the XX limit shows the strongest pinning: its boundary remains almost horizontal around $h_{\rm SVM}-h_c=0$. This behavior is consistent with the sharper kernel structure of the XX limit, where the additional symmetry and magnetization-sector structure make the fidelity geometry particularly strong. The Ising and XY cases also remain well separated from the midpoint response, although their boundary shifts are slightly larger than in the XX case.

The BKT case is more subtle, as expected from the smooth finite-size fidelity feature. In this case, the hard-margin SVM gives a useful reference, but its boundary follows the shifted midpoint more closely than in the XY-family cases. This behavior is natural for the BKT transition, where the fidelity landscape is broader and the finite-size feature is less sharply localized.  We therefore repeat the shifted-window test using a cost-sensitive soft-margin SVM. In the soft-margin formulation the hard-margin constraints of Sec.~\ref{SVM_fid} are relaxed by slack variables, whose sum is penalised in the objective function by a regularisation parameter $C$: a large $C$ penalises margin violations strongly and recovers the hard-margin limit, whereas a small $C$ permits a wider margin at the price of some misclassified training points. Cost-sensitive means that this penalty is assigned separately to the two classes, $C_L$ to the samples of the left window ($y_i=-1$) and $C_R$ to those of the right window ($y_i=+1$), so that the ratio $C_R/C_L$ controls how asymmetrically the two phases are treated. The threshold $T$ generalises the decision rule of Eq.~\eqref{eq:DF}: the boundary is defined by $d(\Delta)=T$ rather than by the zero crossing $d(\Delta)=0$, which displaces it toward one of the two phases. The values $C_L=0.1$, $C_R/C_L=1.122$ and $T=0.25$ are fixed once for the unshifted configuration and then kept unchanged for every window shift. This is therefore a stability test after a single calibration, rather than an independent determination of the BKT point. The mean training accuracy changes only from about $0.981$ in the hard-margin case to about $0.976$ in the soft-margin case, while the extracted boundary becomes less sensitive to the displacement of the training windows.

\begin{figure}[htbp]
    \centering
    \includegraphics[width=\linewidth]{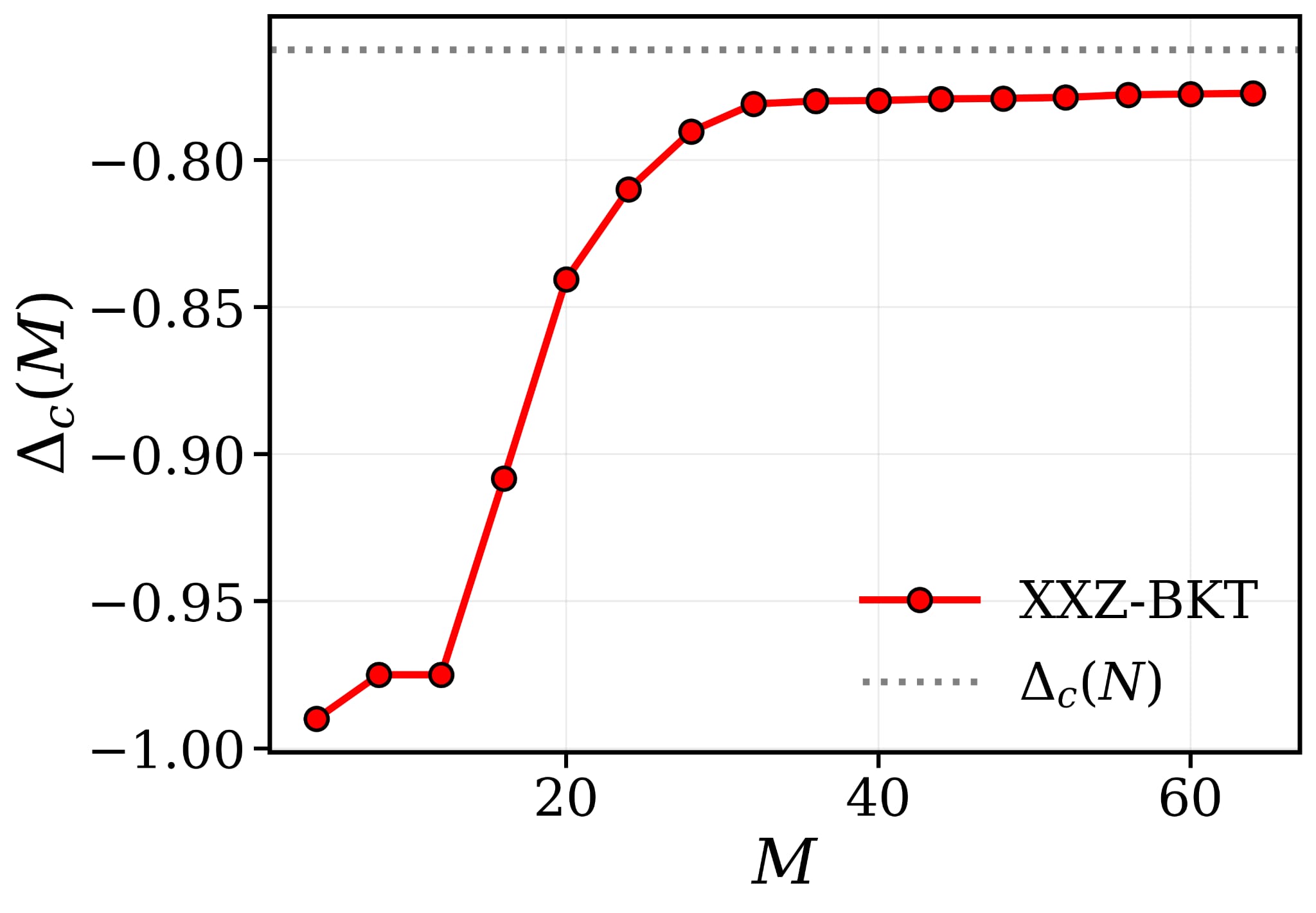}
    \caption{Saturation of the BKT SVM-estimate with the number of training points at fixed system size $N=100$. The horizontal axis shows the total number of training points $M$, while the vertical axis shows the SVM estimate $\Delta_c(M)$. The red curve corresponds to the XXZ-BKT soft-margin SVM, and the gray dotted line marks the finite-size BKT reference point $\Delta_c(N)$. The estimate approaches a stable value as $M$ is increased, showing that the BKT boundary is robust once the training windows are sufficiently sampled.}
    \label{fig:bkt_M_saturation}
\end{figure}

 A second, complementary check is to keep the BKT training windows fixed and increase the number of training points inside them. If the SVM estimate is meaningful, it should not keep changing indefinitely as more points are added instead, it should approach a stable value once the windows are sufficiently sampled. We therefore compute the BKT estimate as a function of the total number of training points $M$ and denote the corresponding SVM estimate by $\Delta_c(M)$. The result is shown in Fig.~\ref{fig:bkt_M_saturation}. As $M$ increases, $\Delta_c(M)$ approaches a stable value close to the finite-size BKT reference $\Delta_c(N)$, providing an additional check that the BKT boundary is not a consequence of using only a small training set.

 We stress that, for the BKT transition, the robustness of the SVM boundary must be assessed differently from the second-order case. For a continuous transition with a fixed thermodynamic critical point, one can hold the training windows fixed and study the finite-size scaling of the SVM estimate directly, as we do for the XY family in Appendix~\ref{app:bogoliubov_fidelity} (Fig.~\ref{fig:FS}). This procedure is not well defined for the BKT boundary, because the fidelity feature that anchors the training windows itself drifts logarithmically with $N$, as shown in Fig.~\ref{fig:xxz_bkt}: windows fixed at the feature location for one system size no longer bracket the transition at another, so a fixed-window $\Delta_{\rm SVM}(N)$ curve would not compare like with like. For this reason, the robustness of the BKT boundary is established here through the two complementary tests already presented, rather than through a fixed-window scaling curve. The shifted-window analysis of Fig.~\ref{fig:sv_midpoint_sensitivity} shows that the soft-margin boundary is not dictated by the placement of the training windows, and the saturation with $M$ in Fig.~\ref{fig:bkt_M_saturation} shows that it is not an artifact of a small training set. Together with the finite-size drift of the underlying fidelity feature up to $N=480$, which extrapolates to $\Delta_c\approx-0.52$ in agreement with the thermodynamic value $\Delta_c=-0.5$ (Fig.~\ref{fig:xxz_bkt}), these establish that the finite-$N$ SVM boundary is a robust pseudo-critical estimator subject to the standard BKT logarithmic drift, rather than a consequence of the chosen training configuration.

\FloatBarrier
\section{Conclusion and Outlook}
We have established a direct link between many-body symmetries and the measurement overhead of quantum kernel estimation, revealing that continuous symmetries systematically increase shot requirements through kernel concentration. 

The symmetry-resource hierarchy $\mathbb{Z}_2$ (Ising/XY) $\ll$ U(1) (XX) $<$ SU(2)-approaching (XXZ) emerges, as shown both analytically and numerically, persisting beyond finite-size effects. It allowed us to tackle the phase transition of finite order with good accuracy when benchmarked with ground-truth values. 
Unlike the preceding works, e.g.~\cite{SanchoLorente2022PRA}, the present work does not focus solely on explaining why fidelity kernels identify second-order phase transitions. 
Rather, it thoroughly deals with resource scaling on quantum hardware for systems of different symmetry, and in parallel, it extends the classification capability of the learning strategy. 
We quantified the quantum resources required to estimate fidelity kernels on finite quantum processors, a necessary step toward realistic implementations of quantum-kernel learning algorithms.
Our results should therefore be interpreted not as an assessment of whether fidelity kernels detect quantum phase transitions, 
which is already well established, but as an assessment of when they remain experimentally implementable on quantum hardware.

The following three implications are given. First, symmetry exploitation in quantum-informed ML is double-edged: while continuous symmetries simplify ground states analytically, they paradoxically degrade kernel discriminability on NISQ hardware. Second, our concentration-avoidance bound $S_\text{CA} \propto 1/k_\text{repr}^2$, as in Eq.~\eqref{eq:SCA-main}, and provides a practical diagnostic for when fidelity kernels fail, favoring discrete-symmetry models for early quantum advantage. Third, the observed non-monotonic shot scaling (minimum $\sim N=30$) suggests optimal system sizes for balancing distinguishability against concentration.

Looking ahead, our paper paves the way for the extensions to the analysis of systems in higher-dimensional lattices for which analytical methods are not available. 
Our symmetry-resource framework establishes a new resource assessment for investigating critical systems on digital quantum simulators, 
in the spirit of physics-informed quantum algorithms.
Our shot bounds enable precise resource accounting for fidelity-kernel estimation on near-term QPUs, 
prioritizing $\mathbb{Z}_2$-symmetric models (Ising/XY) over U(1)/SU(2) cases for scalable phase recognition. 
Extensions to higher-dimensional lattices and physics-informed kernel mitigation, such as regularization, adaptive collocation sampling, 
will unlock digital quantum simulation of regimes fundamentally inaccessible to commonly employed analytical tools, 
ultimately allowing us to explore new phases of matter on early fault-tolerant hardware. 

Fidelity susceptibility is one of the most successful geometric probes of quantum phase transitions 
and has recently become the subject of dedicated quantum algorithms capable of estimating it with Heisenberg-limited precision  \cite{qalgo_fidsus}. 
This development highlights that fidelity-based quantities are increasingly regarded as fundamental computational primitives 
for quantum simulation rather than merely theoretical diagnostics. Our work complements this direction. 
Whereas fidelity susceptibility characterizes local sensitivity of the ground state, fidelity-kernel learning exploits global pairwise overlaps between many-body states. 
The present work addresses the resource requirements associated with estimating these overlaps and demonstrates how their complexity depends on the symmetry of the underlying quantum system.


\section*{Acknowledgments}
This work has received funding from the 
European Union — Next Generation EU 
through PNRR MUR Project No.~CN00000013, 
“Italian National Centre on HPC, Big Data and Quantum Computing” 
and PNRR MUR Project No.~PE0000023-NQSTI, “National Quantum Science and Technology Institute"; 
and INFN through the project “QUANTUM.”
AM acknowledges the IRA Programme, project no.~FENG.02.01-IP.05-0006/23, financed by the FENG program 2021-2027, Priority FENG.02, Measure FENG.02.01., with the support of
the FNP.
\section*{Data and code availability}
All results reported in this work are based on \emph{synthetic} data generated numerically. No experimental datasets were used.
The code used to generate the synthetic states, compute the labels and features, train the machine-learning models, and reproduce all
figures and numerical results will be shared by the corresponding author upon reasonable request.
\bibliography{bibliography_1}

\appendix

\section{Bogoliubov-angle fidelity for the transverse-field XY chain} \label{app:bogoliubov_fidelity}

For completeness, we briefly review the standard diagonalization of the transverse-field XY chain and derive the closed-form expression for the ground-state fidelity used in the main text.

We consider the transverse-field XY model with periodic boundary conditions,
\begin{align}
H(\gamma,h)= -&\sum_{j=1}^N\!\left(\frac{1+\gamma}{2}\,\sigma_j^x\sigma_{j+1}^x
+\frac{1-\gamma}{2}\,\sigma_j^y\sigma_{j+1}^y\right)\nonumber\\
-& h\sum_{j=1}^N \sigma_j^z ,
\label{eq:app_H_XY}
\end{align}

Under the Jordan--Wigner transformation, the spin Hamiltonian becomes quadratic in fermionic operators up to a boundary term whose sign depends on the fermion-parity sector. In the finite-size analysis considered here, we work in the even-parity sector, which leads to antiperiodic boundary conditions for the fermions. The transformation reads
\begin{align}
\sigma_j^+ &= c_j^\dagger \prod_{\ell=1}^{j-1}(-\sigma_\ell^z),\qquad
\sigma_j^- = \left(\prod_{\ell=1}^{j-1}(-\sigma_\ell^z)\right)c_j, \nonumber\\
\sigma_j^z &= 1-2c_j^\dagger c_j ,
\label{eq:app_JW}
\end{align}
Eq.~\eqref{eq:app_H_XY} becomes (up to an additive constant)
\begin{align}
H(\gamma,h)=&-\sum_{j=1}^N\!\Big(c_j^\dagger c_{j+1}+c_{j+1}^\dagger c_j \label{eq:app_H_JW}\\
&\qquad\quad +\gamma\,c_j^\dagger c_{j+1}^\dagger +\gamma\,c_{j+1} c_j\Big)
+2h\sum_{j=1}^N c_j^\dagger c_j .
\nonumber
\end{align}
Let $N_f=\sum_j c_j^\dagger c_j$ be the fermion-number operator and $\Pi=(-1)^{N_f}$ the fermion-parity operator. Restricting to the even-parity sector, $\Pi=+1$, implies antiperiodic boundary conditions $c_{N+1}=-c_1$, and therefore the allowed momenta are
\begin{equation}
q=\frac{(2m+1)\pi}{N},\qquad m=0,1,\dots,\frac{N}{2}-1 .
\label{eq:app_k}
\end{equation}
Using $c_j=\frac{1}{\sqrt{N}}\sum_q e^{iqj}c_q$, the Hamiltonian decouples into independent $(q,-q)$ sectors,
\begin{equation}
H(\gamma,h)=\sum_{q>0} H_q + \mathrm{const},
\label{eq:app_Hsum}
\end{equation}
with
\begin{equation}
H_q=
2\epsilon_q\big(c_q^\dagger c_q + c_{-q}^\dagger c_{-q}\big)
+2i\Lambda_q\big(c_q^\dagger c_{-q}^\dagger - c_{-q}c_q\big),
\label{eq:app_Hk}
\end{equation}
where
\begin{equation}
\epsilon_q := h-\cos q,\qquad \Lambda_q := \gamma\sin q .
\label{eq:app_xi_Delta}
\end{equation}
The first term is diagonal in the occupation basis whereas the second term is not yet diagonal and this motivates the Bogoliubov transformation 
\begin{align}
    \eta_q=u_qc_q-iv_qc_{-q}^\dagger
    \quad
     \eta_{-q}=u_qc_{-q}+iv_qc_{q}^\dagger
\end{align}
    
A convenient choice is $u_q=\cos\theta_q$, $v_q=\sin\theta_q$ for which $H_q$ is diagonal provided
\begin{equation}
\tan\!\big(2\theta_q(h)\big)=\frac{\Lambda_q}{\epsilon_q}
=\frac{\gamma\sin q}{\,h-\cos q\,}.
\label{eq:app_tan2theta}
\end{equation}
With this choice,
\begin{equation}
H_q = 2 E_q \big( \eta_q^\dagger\eta_q+\eta_{-q}^\dagger\eta_{-q}-1 \big),
\label{eq:Hk_diag}
\end{equation}
Where  dispersion is
\begin{equation}
E_q=\sqrt{\epsilon_q^2+\Lambda_q^2}=\sqrt{(h-\cos q)^2+\gamma^2\sin^2 q}.
\label{eq:app_disp}
\end{equation}

This yields a closed-form expression for the Bogoliubov angle. A convenient explicit choice consistent with Eq.~\eqref{eq:app_tan2theta} is
\begin{equation}
\theta_q(h)=\frac{1}{2}\arg\!\left[\epsilon_q(h)+i\Lambda_q\right]
\label{eq:app_theta_arctan}
\end{equation}
The two-argument form fixes the branch so that $\theta_q\to0$ for $h>\cos q$ and $\theta_q\to\pi/2$ for $h<\cos q$, that is, so that $v_q=\sin\theta_q$ reproduces the occupation of the mode in the ground state.
Because the ground state factorizes over $q>0$, the fidelity between two ground states at fields
$h_a$ and $h_b$ (same $\gamma$) factorizes mode by mode as
\begin{equation}
F(h_a,h_b)
:=\left|\braket{\psi(h_a)}{\psi(h_b)}\right|^2
=\prod_{q>0}\cos^2\!\Big(\theta_q(h_b)-\theta_q(h_a)\Big).
\label{eq:app_fidelity_product}
\end{equation}
In Eq.~\eqref{eq:app_fidelity_product}, $q$ runs over the even-parity momenta in Eq.~\eqref{eq:app_k}, and $\theta_q(h)$ is given by Eq.~\eqref{eq:app_theta_arctan}. Equation~\eqref{eq:app_fidelity_product} is the analytical expression used in the main text to construct both the global fidelity kernel in Eq.~\eqref{eq:global} and the fidelity-per-site kernel in Eq.~\eqref{eq:fid_per_site}.

\begin{figure}[H]
    \centering
    \includegraphics[width=0.9\linewidth]{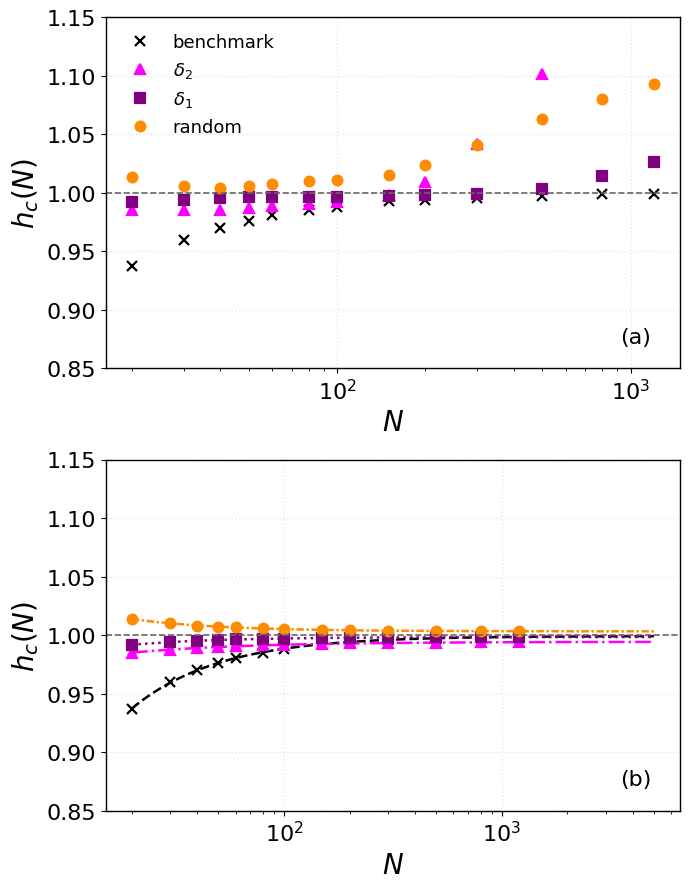}
    \caption{Finite-size scaling of the pseudo-critical field in the XY chain ($\gamma=0.5$). For each system size $N$ (log scale on the horizontal axis), we extract a pseudo-critical point $h_c(N)$ using: (i) a training-free benchmark (black $\times$), with $h_{\mathrm{ref}}=1.75$ and (ii) an SVM decision-boundary estimate $d(h)=0$ trained on $M=16$ labeled points (8 per phase) with three sampling strategies: $\delta_1=[0.85,0.90]\cup[1.10,1.15]$ (purple squares), $\delta_2=[0.60,0.70]\cup[1.60,1.70]$ (magenta triangles), and random uniform sampling in $[0.25,1.75]$ labeled according to the sign of $h-1$ (yellow circles). \emph{Top panel:} SVM trained with the global fidelity kernel, Eq.~\eqref{eq:global}. \emph{Bottom panel:} SVM trained with the fidelity-per-site kernel, Eq.~\eqref{eq:fid_per_site}. The horizontal dotted line marks the thermodynamic critical value $h_c=1$. The dashed and dotted curves are fits  showing the convergence of $h_c(N)$ with increasing $N$.}
    \label{fig:FS}
\end{figure}

We further use this closed-form fidelity to compare SVM-based pseudo-critical estimates with a training-free benchmark. For the latter, we define
\begin{equation}
h_c^{(\mathrm{bench})}(N)
=\arg\max_{h}\Big|\partial_h\,\mathcal{S}(h,h_{\mathrm{ref}})\Big|,
\end{equation}
where $\mathcal{S}(h,h_{\mathrm{ref}})$ denotes a similarity measure built either from the global fidelity, Eq.~\eqref{eq:global}, or from the fidelity per site, Eq.~\eqref{eq:fid_per_site}, with $h_{\mathrm{ref}}$ a fixed reference field in the paramagnetic regime far from the critical point. To extrapolate the finite-size pseudo-critical points, we use the empirical drift form~\cite{fisher1972scaling}
\begin{equation}
h_c(N)=h_c+\mathcal{A}\,N^{-p},
\end{equation}
where $p$ denotes an effective finite-size drift exponent which, together with $\mathcal{A}$, is treated as a fitting parameter. The resulting fits are shown in Fig.~\ref{fig:FS} and summarized in Table~\ref{tab:drift_fits_hcN_svm}.

\begin{table*}[htbp!]
\centering
\begin{ruledtabular}
\begin{tabular}{@{}lccc@{}}
Method & $h_c$ & $p$ & $\mathcal{A}$ \\\\
\hline
benchmark & $0.999339\pm0.000544$ & $0.945\pm0.126$ & $-1.458\times 10^{0}\pm9.092\times 10^{-1}$ \\
$\delta_2$ & $0.994163\pm0.000020$ & $1.170\pm0.022$ & $-1.304\times 10^{-1}\pm8.883\times 10^{-3}$ \\
$\delta_1$ & $0.998629\pm0.000011$ & $1.469\pm0.020$ & $-3.834\times 10^{-2}\pm1.451\times 10^{-3}$ \\
random & $1.002729\pm0.000009$ & $1.292\pm0.010$ & $8.741\times 10^{-2}\pm2.203\times 10^{-3}$ \\
\end{tabular}
\end{ruledtabular}
\caption{Finite-size scaling fit of the empirical drift form $h_c(N)=h_c+\mathcal{A}\,N^{-p}$ for the pseudo-critical points extracted from the benchmark and SVM-based estimators. Reported values are best-fit parameters, and the quoted uncertainties are one standard deviation, obtained as the square root of the diagonal elements of the fit covariance matrix.}\label{tab:drift_fits_hcN_svm}
\end{table*}

\end{document}